\documentclass[article]{aa}

\usepackage{graphicx}
\usepackage{txfonts}
\usepackage[colorlinks=true, allcolors = blue]{hyperref}
\usepackage{subfigure}
\usepackage{threeparttable}

\usepackage{xcolor}

\begin{document}

   \title{Magnetically driven winds from accretion disks in post-asymptotic giant branch binaries}


   \author{Olivier Verhamme
          \inst{1}, Jacques Kluska \inst{1},
          Jonathan Ferreira \inst{2}, Dylan Bollen \inst{1,3}, Toon De Prins \inst{1}, Devika Kamath \inst{3,4}, Hans Van Winckel \inst{1}
          }

   \institute{Institute of Astronomy, KU Leuven, Celestijnenlaan 200D, B-3001 Leuven, Belgium \\
              \email{Olivier.Verhamme@kuleuven.be}
         \and
   Univ. Grenoble Alps, CNRS, IPAG, F-38000 Grenoble, France
          \and
    Department of Physics \& Astronomy, School of Mathematical and Physical Sciences, Macquarie University,
                  Sydney, NSW 2109, Australia
                  \and
    Astronomy, Astrophysics and Astrophotonics Research Centre, Macquarie University, Sydney, NSW 2109, Australia
            }

   \date{Received Oktober 2023; Accepted December 2023}
\titlerunning{Jets in post-AGB binaries}
\authorrunning{Verhamme et al.}

  \abstract
   {Jets are commonly detected in post-asymptotic giant branch (post-AGB) binaries and originate from an accretion process onto the companion of the post-AGB primary. These jets are revealed by high-resolution spectral time series.}
   {This paper is part of a series.
   In this work, we move away from our previous parametric modelling and include a self-similar wind model that allows the physical properties of post-AGB binaries to be characterised.
   This model describes magnetically driven jets from a thin accretion disk threaded by a large-scale, near equipartition vertical field.}
   {We expanded our methodology in order to simulate the high-resolution dynamic spectra coming from the obscuration of the primary by the jets launched by the companion. We present the framework to exploit the self-similar jet models for post-AGB binaries. We performed a parameter study to investigate the impact of different parameters (inclination, accretion rate, inner and outer launching radius) on the synthetic spectra.}
   {We successfully included the physical jet models into our framework. The synthetic spectra have a very similar orbital phase coverage and absorption strengths as the observational data. The magnetohydrodynamic (MHD) jet models provide a good representation of the actual jet creation process in these evolved binaries. Challenges remain, however, as the needed high-accretion rate would induce accretion disks that are too hot in comparison to the data. Moreover, the rotational signature of the models is not detected in the observations. In future research, we will explore models with a higher disk ejection efficiency and even lower magnetisation in order to solve some of the remaining discrepancies between the observed and synthetic dynamic spectra.}
   {}

   \keywords{Stars: AGB and post-AGB – Stars --
            binaries: spectroscopic --
            Stars: mass-loss --
            Stars: jets  --
            Accretion, accretion disks
               }

   \maketitle

%

\section{Introduction}

\begin{figure}
    \centering
    \includegraphics[width=0.45\textwidth]{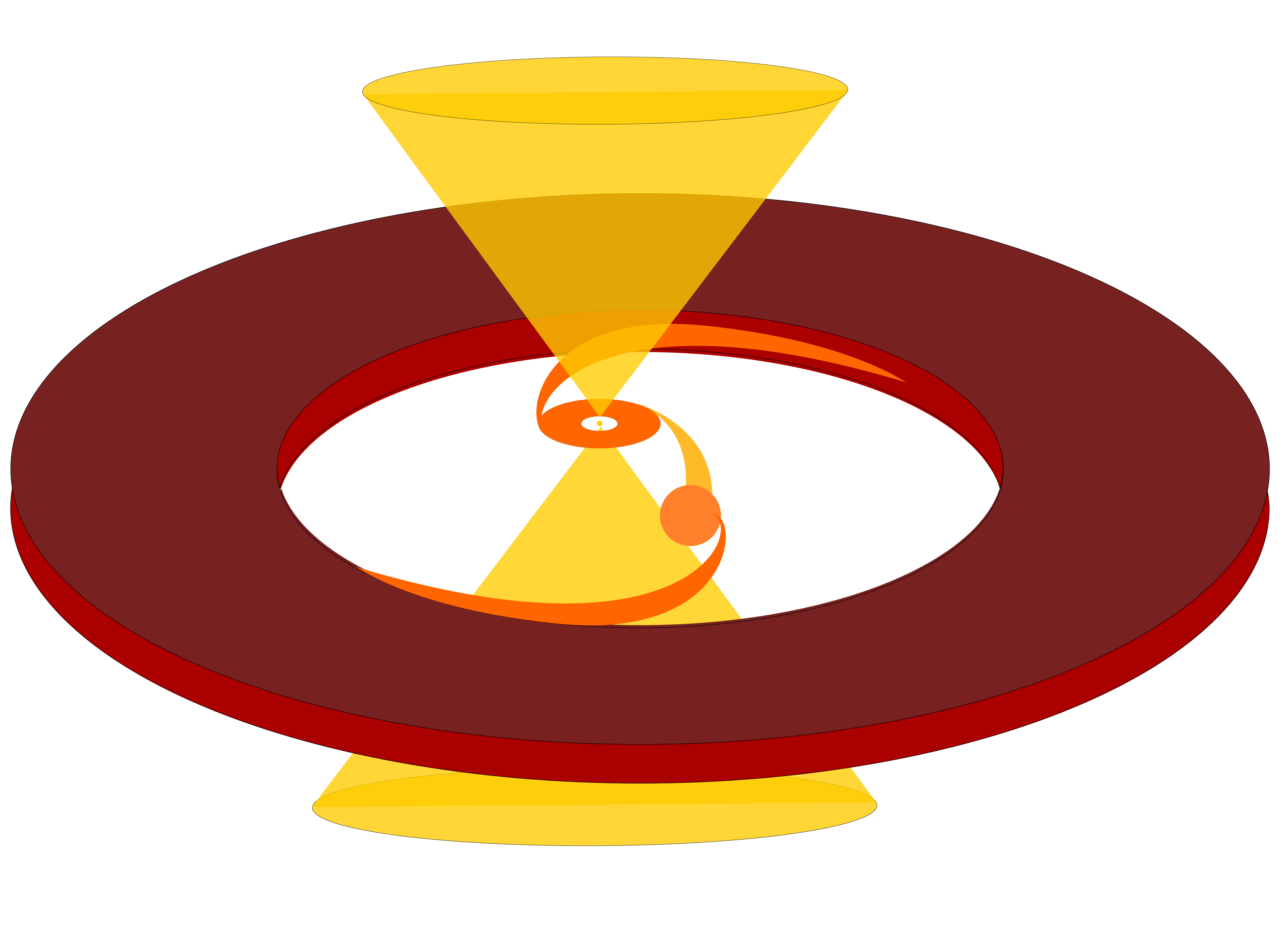}
    \caption{Conceptual figure of post-AGB binaries. Figure is from \cite{bollen_structure_2022}.
    \label{fig:blocks}}
\end{figure}

In this paper, we discuss disk winds launched by main-sequence companions in a binary with a post-asymptotic giant branch (post-AGB) primary \citep{van_winckel_post-agb_2003}.
We use the term `post-AGB star' to refer to a star that has just evolved off the AGB branch, meaning the star is contracting and heating at a constant luminosity up to a value of $10^{5} L_{\odot}$, with mass-loss rate estimates between $10^{-9}-10^{-7} \frac{M_{\odot}}{yr}$ \citep{oomen_modelling_2019, van_winckel_binary_2018, van_winckel_post-agb_2003}.
The most important building blocks (see Figure \ref{fig:blocks}) are the evolved primary and its main-sequence companion, which are in an orbit ranging from a couple hundred to a few thousand days and are surrounded by a stable circumbinary disk of gas and dust.
A catalogue of Galactic binary post-AGB systems was recently published by \cite{kluska_population_2022}.




\begin{figure}
    \centering
    \includegraphics[width=0.45\textwidth]{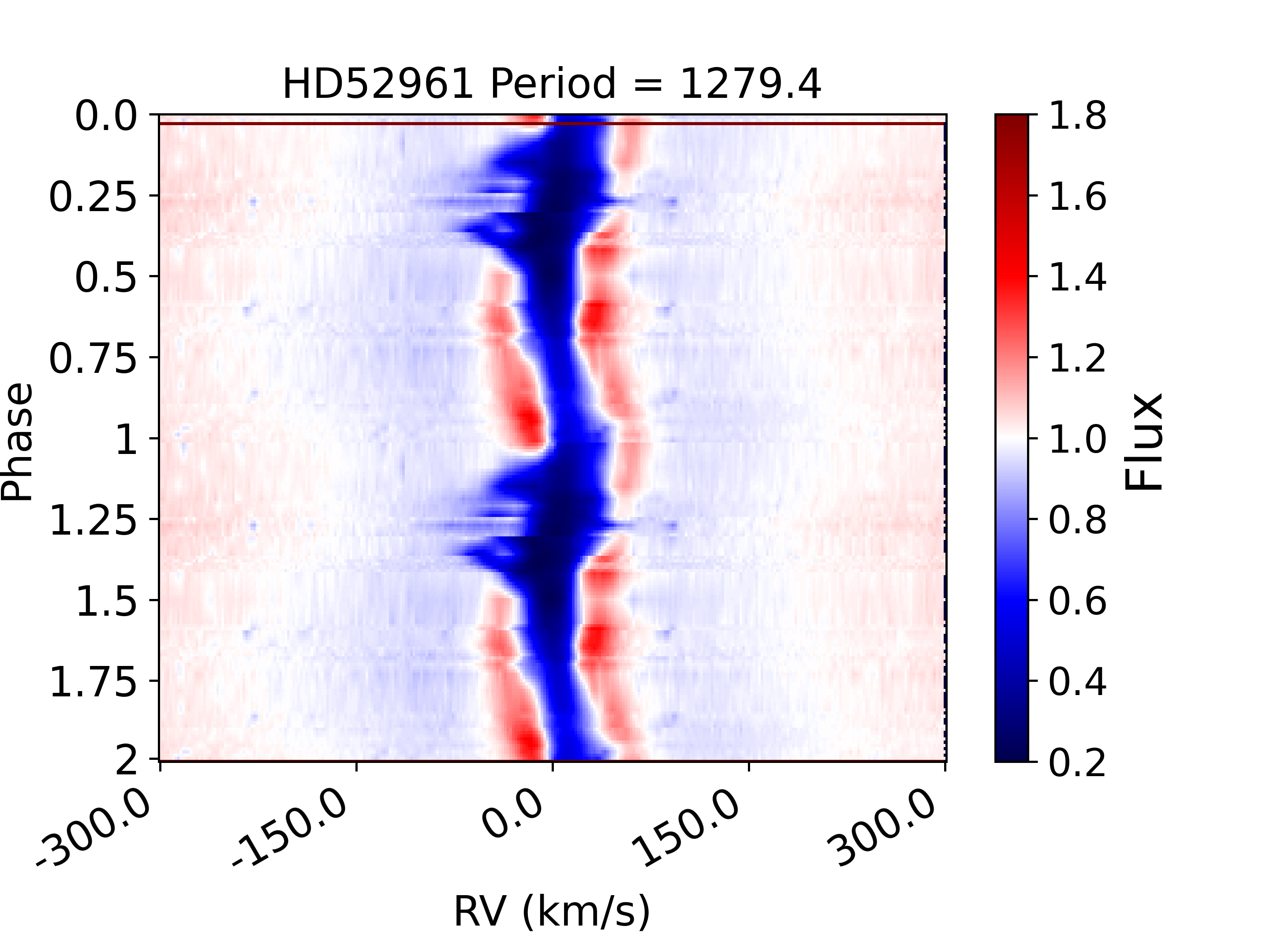}
    \includegraphics[width=0.45\textwidth]{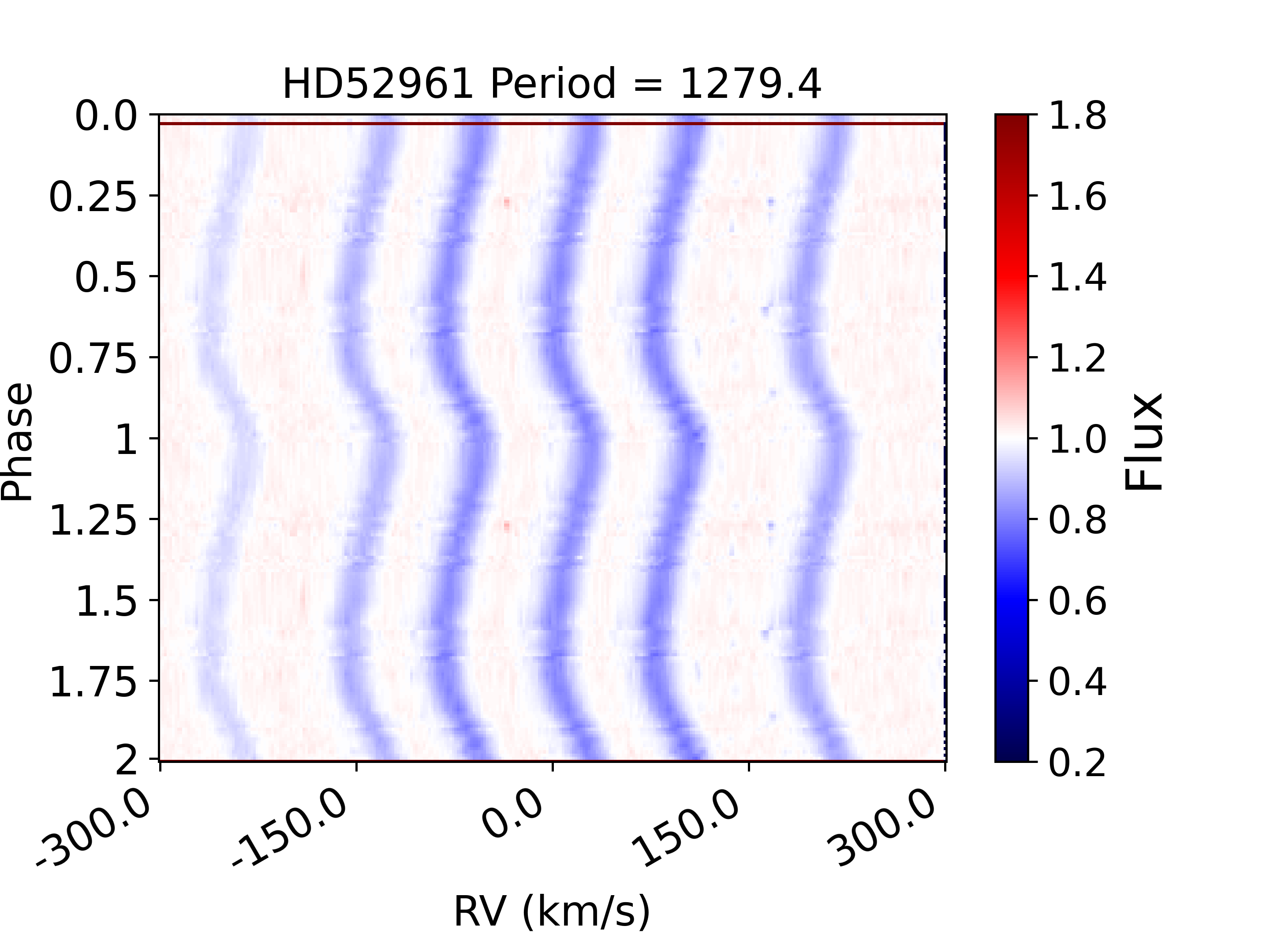}
    \caption{Observable influence of the orbit and jet on the spectra. The top panel gives the dynamic observed spectra of the $H_{\alpha}$ profile of the post-AGB star HD52961 folded onto the orbital period. The initial phase is random. The lower panel gives the same but for the C~I multiplet around 7115$\AA$. The orbital phase is shown twice to guide the eye.}
    \label{fig:spectra}
\end{figure}

The luminosity of the primary dominates the optical spectrum, but the thermal contribution of the accretion disk around the companion can be resolved with near-infrared interferometry \citep{hillen_imaging_2016, kluska_perturbed_2018}.
The observation of an accretion disk strengthens the belief that the bipolar outflow is launched from the circum-companion accretion disk.
This outflow is not resolved, but its presence can be deduced indirectly by time-resolved high-resolution optical spectra.
The outflow is seen as a high-velocity and phase-dependent absorption feature in the Balmer lines, which also follow the movement of the secondary companion.
This absorption component appears around a superior conjunction \citep{van_winckel_post-agb_2007,gorlova_time-resolved_2012,bollen_jet_2017}.
When the companion moves in front of the primary, one of the cones of the jet gradually enters the line of sight (LOS) towards the primary.
This creates an absorption component as continuum photons from the background post-AGB primary are selectively removed from the LOS.
We illustrate this in Figure~\ref{fig:spectra}.
In the figure, we show our time series of high-resolution optical spectra as obtained by our HERMES spectrograph \citep{raskin_hermes_2011, raskin_2011} mounted on the Mercator telescope at the Roque de los Muchachos observatory in Spain.
We folded the spectra around H$_{\alpha}$ on the orbital period and display the spectra in velocity space.
At the superior conjunction, one can see the outflow in absorption when the bicone is in the LOS towards the primary.

The observed collimated outflows can be described as extended disk winds launched from this accretion disk \citep{blandford_hydromagnetic_1982, ferreira_fan-shaped_2013}.
This type of magneto-centrifugal outflow is formed when the magnetic field lines are threading the disk at an angle acute enough for the material to be transported by the centrifugal force over the field lines away from the accretion disk.
This type of disk wind can occur even at low magnetisation \citep{jacquemin-ide_magnetically_2019, jacquemin-ide_magnetic_2020}.


This paper is a continuation of a series of papers \citep{bollen_jet_2017, bollen_spatio-kinematic_2019, bollen_determining_2020, bollen_structure_2022} describing both the observational effort as well as the modelling efforts to exploit the spectral time series in order to model the jets and to use these systems to study the launching physics of jets.
In our previous papers, we focussed on developing a geometric parametric model to fit the observed time series.
This was followed by a line-tracing radiative transfer module for quantifying the total mass loss in the jets.
The methodology is outlined in \cite{bollen_structure_2022} and was applied to a total of 16 objects.
The jet model used in these papers is purely geometric, with a jet axis and an angle-dependent density and
velocity law.
The results of these studies have also shown that high mass-ejection rates are needed (reaching values up to $10^{-4}\frac{\dot{M}_{\odot}}{yr}$) and hence so are high accretion rates onto the accretion disk from the circumbinary disk.
To further study these surprisingly large mass accretion rates, further investigation is necessary.

To that end, we use the same setup in this work but move away from the geometric models and include self-similar physical jet models.
These were developed in the framework jets in YSOs \citep{jacquemin-ide_magnetically_2019}.
While the long-term goal is to reproduce the data quantitatively, in this work, we present the framework to use these self-similar jet models for post-AGB binary systems.
We qualitatively compare the results with the observables of one target and observe the influence of the different model parameters without fine-tuning the model to reproduce the data.
We organised the paper as
follows: In section \ref{Spatio-kinematic-model}, we give a short overview of the methodology employed up to now using geometric biconal models.
Next, we focus on the properties of the self-similar jet models and how these are implemented (Sect \ref{Self-con model}).
We perform a parameter study of the physical parameters and focus on their impact on the observables, namely, the time series of high-resolution $H_{\alpha}$ profiles.
We analyse our parameter study in Section \ref{sec:results}.
We conclude our paper by highlighting our most important findings and by isolating some avenues for future research.

\section{Initial framework: The parametric model}\label{Spatio-kinematic-model}

We give a brief description of the methodology used in previous studies
\citep[][and references therein]{bollen_structure_2022}.
The parametric model is a purely geometric model of the jet, and we use it to generate synthetic dynamical spectra that can be compared to the observed spectra.
The first step in this approach is to define the initial spectrum that will serve as the basis on which the absorption from the jet model is added (Sect.\,\ref{sec:initialspectra}).
This initial spectrum was also used for the self-similar models in this work.
Then, the parameters of the jet model are constrained as described in Sect.\,\ref{sec:JetModel}.
Finally, the absorption profile is obtained by performing a radiative transfer step (Sect.\,\ref{sec:RadiativeTransfer}).



\subsection{Initial spectra}

\label{sec:initialspectra}


To estimate the impact of an obscuring jet bicone, it is necessary to determine the unobscured spectrum at all orbital phases.
This unobscured spectrum is assumed to be the spectrum observed with no circumsecondary jet and serves as a background spectrum from which the jet obscuration is computed.
One of the issues in determining the background spectra is that the
unabsorbed base spectra of the H-$\alpha$ profile does not correspond
to a purely photospheric spectrum.
Outside the conjunction, the H-$\alpha$ profile shows a double-peaked profile, the origin of which is still unclear.
It is necessary to create a background spectra that takes this into account. In this process, we assume that this background spectrum is orbit independent.

\begin{figure}
    \centering
     \subfigure{\includegraphics[width=0.45\textwidth]{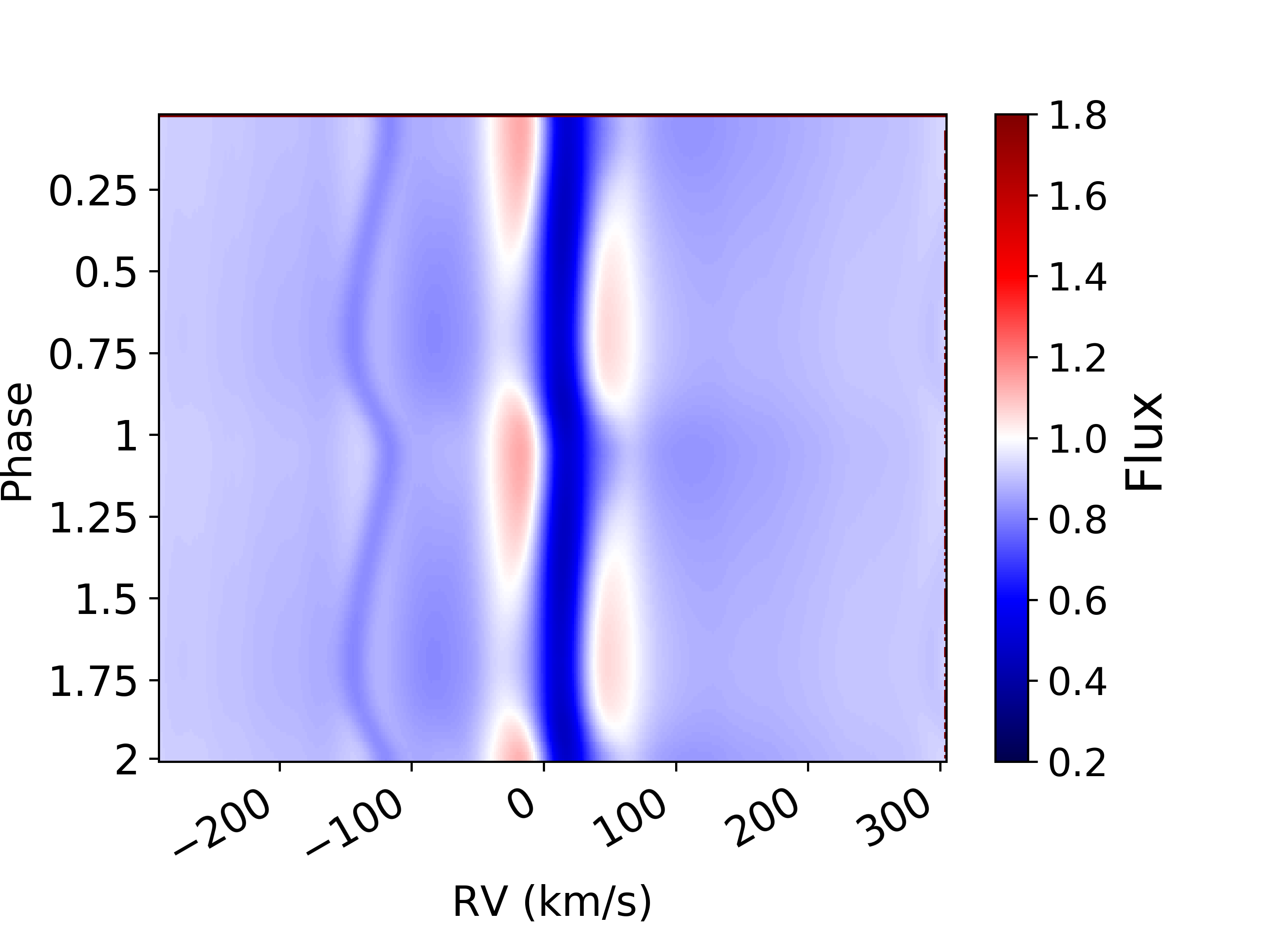}}
   	\hspace{0.025\textwidth}
    \subfigure{\includegraphics[width=0.45\textwidth]{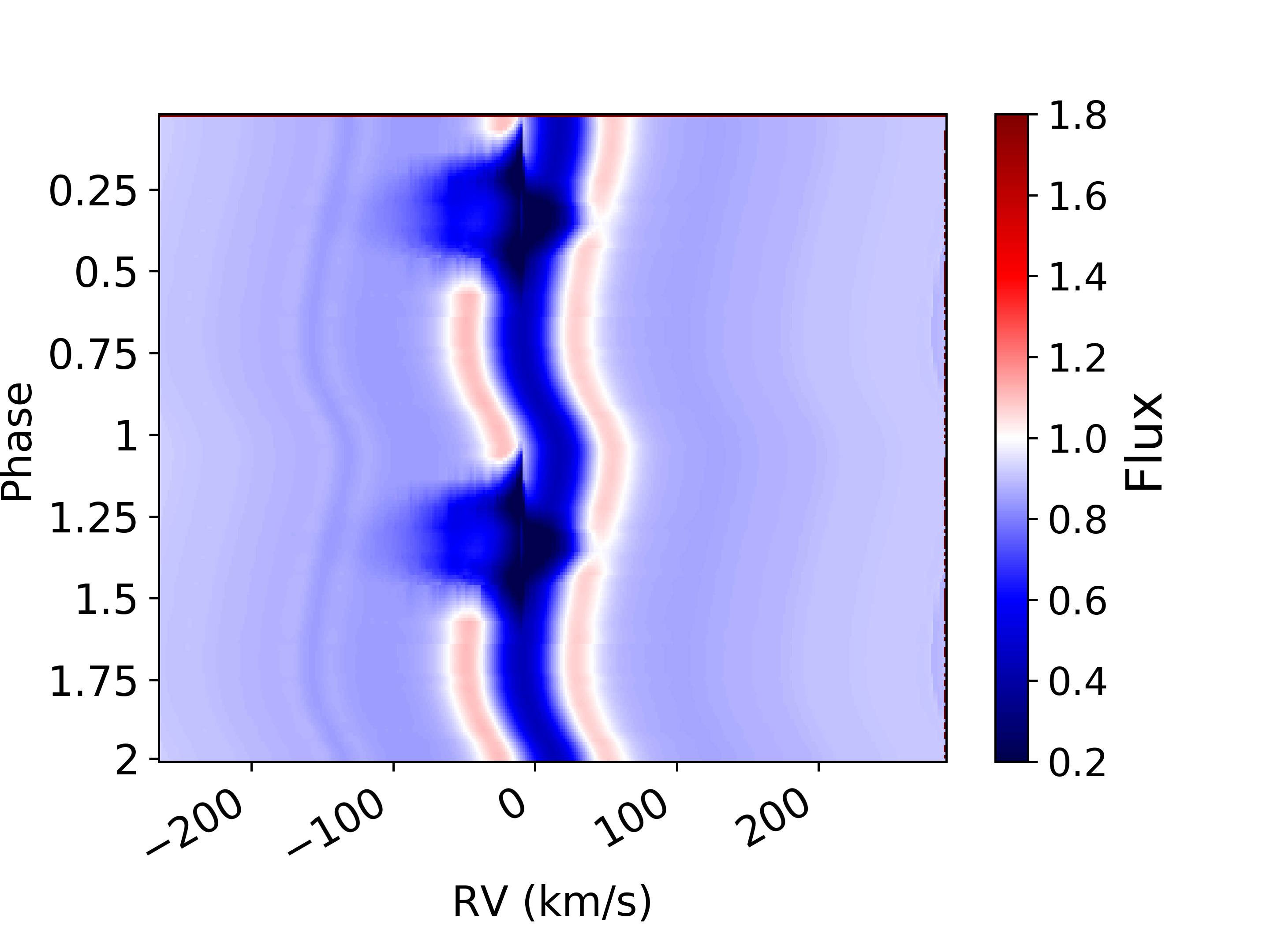}}
   	\hspace{0.025\textwidth}
    \caption{Modelling effect on the dynamic spectra: The top panel shows the initial H-$\alpha$ line-profile of HD 52961 which serves as background spectrum. The lower panel shows the synthetic dynamic spectra of the H-$\alpha$ line using the stellar disk wind parametric model.}
    \label{fig:Background}
\end{figure}

An example of how the dynamic background spectra appear without the absorption of the jet is given in Figure~\ref{fig:Background}.
To determine this background, we used the average of observations in phases where no jet absorption was seen and assumed that this is a valid background spectrum also at phases of conjunction.
To use these spectra correctly, it is important to understand the orbital motion of the system.
The wavelength shift of the spectra reveals the radial velocity of the post-AGB star around the centre of mass.
We used the orbital parameters of \cite{oomen_orbital_2018} given in Table \ref{tab:HD_Orbital_Param}.
Three of the spectra outside the conjunction contribute to a mean spectrum that serves as a new base spectrum.
This spectrum is then used to create a synthetic dynamic spectrum by reintroducing the correct wavelength shift for the specific orbital phase of the post-AGB star.

\begin{table*}
    \centering
    \caption{Orbital parameter and best spatiokinematic fit.}
    \begin{tabular}{c|c||c|c}

         &   orbital parameters & & Parametric fit \\
         \hline
        Period [Days] & 1279.4$\pm$ 0.23&  $i [^\circ]$& 69.88 \\
        Eccentricity & $0.22 \pm 0.00$ & $\theta_{\text{out}} [^\circ]$ &  49.56\\
        $T_0$ [Days] & $2456309 \pm 112$ &  $\theta_{\text{in}} [^\circ]$ &  49.31\\
        $\omega$ [$^{\circ}$] & $11 \pm 73$ &  $\theta_{\text{cav}} [^\circ]$ &  25.14\\
        $K_1$ [$\frac{km}{s}$] & $13.264 \pm 0.005$ &  $\phi_{\text{tilt}} [^\circ]$ &  -4.65\\
        $\gamma$ [$\frac{km}{s}$]& $6.1 \pm 0.1$  &  $v_{\text{in}} [\frac{km}{s}]$& 158\\
        semi-major axis (a) [$R_{\odot}$]& 586&$v_{\text{out}} [\frac{km}{s}]$& 37\\
         Roche-lobe Radius ($R_L$) [$R_{\odot}$]& 395&$c_{v} $& 0.04\\
         &&$p_{v} $& 2.32\\
         &&$p_{\rho, \text{in}} $& 14.92\\
         &&$p_{\rho, \text{out}} $& -9.89\\
         &&$c_{\tau} $& 2.29\\
         &&$R_{\text{1}} [R_{\odot}]$ & 105.4 \\
         &&$R_{\text{2}} [R_{\odot}]$ & 1.04\\
         \hline
         &&$R_\text{in} [R_{2}]$ & 8.6\\
         &&$R_\text{out} [R_L]$ & 1.00\\
         \hline
    \end{tabular}
    \begin{tablenotes}
        \item \textbf{Notes}: Listed in the table are all orbital parameters of the object HD 52961. Here, `Period' is the time of one revolution in days; `Eccentricity' defines the elongation of the orbit; $T_0$ sets the position of the orbit at phase zero; $\omega$ is the argument of periapsis, which defines the orientation of the ellipse; $K_1$ is the highest radial velocity offset from the median; $\gamma$ is the radial velocity of the mass centre. The semi-major axis and the Roche-lobe radius was calculated assuming an inclination of $60^{\circ}$.\\
    The second column shows the best-fit parameters of HD 52961 found using the parametric fit described in \cite{bollen_structure_2022}.
     The tabulated parameters are: (i) inclination angle of the binary system; ($\theta_{\text{out}}$) jet outer angle; ($\theta_{\text{in}}$) jet inner angle; ($\theta_{\text{cav}}$) jet cavity angle; ($\phi_{\text{tilt}} $) jet tilt; ($v_{\text{in}}$) inner jet velocity; ($v_{\text{out}}$) jet velocity at the jet edges; ($c_{v}$) velocity scaling parameter; ($p_{v} $) exponent for the velocity profile; ($p_{\rho, \text{in}} $, $p_{\rho, \text{out}} $) exponent for the density profile for the outer and inner region; ($c_{\tau} $) optical depth scaling parameter; ( $R_1$) the radius of the post-AGB star. The term $R_2$ is the radius of the secondary.
    Underneath the line, the inner and outer launching radius is shown.
    \end{tablenotes}
    \label{tab:HD_Orbital_Param}
\end{table*}

\subsection{Modelled components}  \label{sec:JetModel}

The model consists of two main components: the post-AGB star itself and the biconical jet centred on the companion.
The post-AGB star is orders of magnitude brighter than its typical main-sequence low-mass companion, so we assumed that this binary partner will not contribute to the resulting spectra of the system.
In optical wavelengths, the same holds true for the circumbinary disk and the secondary accretion disk.
The post-AGB primary and the biconical jet orbit each other with certain binary parameters.
At some orbital phases, a LOS from the surface of the post-AGB star passes through the bicone.
This ray, characterised by the background spectrum, interacts with the material in the jet at the given Doppler velocity, and this forms the basis of this model.

The post-AGB star is represented by a uniform disk perpendicular to the LOS.
This disk is divided into small regions via a Fibonacci grid in which every grid point covers an equal surface area.
We then determined the LOS of every grid point in the disk.

\begin{figure}
    \centering
    \includegraphics[width = 0.45\textwidth]{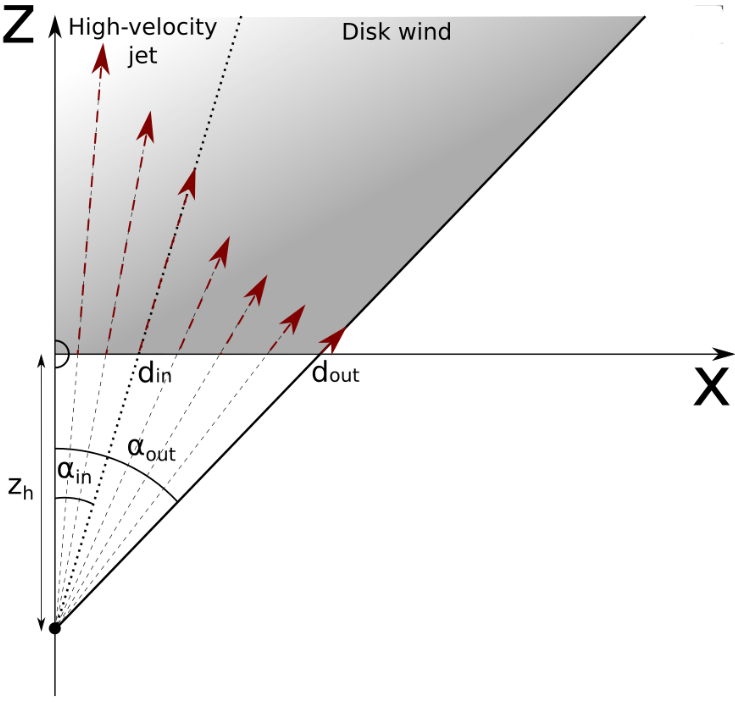}
    \caption{Disk wind model. In the parametric disk wind model, the disk wind is divided into two parts: the inner wind, up to an opening angle of $\alpha_{\text{in}}$, and a disk wind, from $\alpha_{\text{in}}$ up to $\alpha_{\text{out}}$. The position of the point of the cone is set by the combination of the outer launching radius of the disk wind ($d_{\text{out}}$) and  $\alpha_{\text{out}}$.}
    \label{fig:Disk-wind-Figure}
\end{figure}

The disk wind itself is represented by a cone, which is shown in Figure \ref{fig:Disk-wind-Figure}.
This cone has its vertex not located at the secondary star but somewhere under the secondary star, depending on the outer launching radius of the disk wind and the opening angle.
The disk wind has two parts: one central fast part and an outer slower part with a higher density.
A fit angle determines where the fast region stops and the slow region begins. This angle, together with the offset of the cone, corresponds to a certain inner launching radius.

The inner and outer launching radii cannot be chosen freely.
To start, the inner launching radius has to be larger than the stellar radius.
The outer disk radius must be smaller than the Roche-lobe radius of our main-sequence star, as material outside of this lobe will not be bound to the star \citep{eggleton_approximations_1983}:
\begin{align}
    \frac{r_L}{a} = \frac{0.49q^{2/3}}{0.69q^{2/3}+ln(1+q^{1/3})}.
\end{align}
Here, $q$ is the mass ratio between the two stars, and $a$ is the semi-major axis of the primary.
To estimate the radius of the main-sequence companion, we used $R_2 = 1.01 \left( \frac{M_2}{M_{\odot}} \right)^{0.724}$ \citep{demircan_stellar_1991}.


\subsection{Fitting procedure}
\label{sec:Fitting}
For each phase, the global H-alpha profiles were determined using the jet parameters of inclination, opening angle, velocity structure, and scaled density distribution as well as the radius of the post-AGB star and its orbit.
A Markov chain Monte Carlo (MCMC) fitting routine was used to fit these synthetic spectral time series to the observations.
We illustrate this in Fig. \ref{fig:Background} panel b, where we show the simulated dynamic spectrum compared with the top panel of Figure ~\ref{fig:spectra}, which are the observables themselves.
This best fit is based on a disk wind model. The parameters of this best fit are given in the right side of Table \ref{tab:HD_Orbital_Param}.

\subsection{Radiative transfer}
\label{sec:RadiativeTransfer}

To determine the actual mass-loss rates, we needed to compute the absolute densities and assumed temperatures instead of using scaling parameters.
To do so, the parametric model was followed by a radiative transfer module using a temperature and density fit while all other parameters from the parametric fitting were kept constant.
The jet was assumed to be isothermal.
The observables to be reproduced were then the impact of the jet on the Balmer series ($H_\alpha$ to $H_\delta$). The equivalent width of the absorption was used as a tracer of the total opacity.

The source function $S_\nu$ is described by the Planck function $B_\nu$.
As we only needed to find the best fits for two parameters,  a grid of jet temperatures and densities were computed, and each of these models was compared to the data.
This procedure is described in detail in \cite{bollen_determining_2020} and allowed us to determine the total mass loss from the jet while assuming the bicone is symmetric in both lobes.
The found best-fit disk wind mass-ejection rate for the model described in Table \ref{tab:HD_Orbital_Param} is $2.6\cdot 10^{-7} M_{\odot}/yr$, with a 1-$\sigma$ interval of $[1.3 , 21 ]  10^{-7} M_{\odot}/yr$.

\subsection{Overview results}
A full overview of the results obtained for 16 objects can be found in \cite[][and references therein]{bollen_structure_2022}.
The main results to highlight are as follows:

\begin{itemize}
\item The geometric models are versatile and are able to fit a wide variety of time series.

\item The jets are wide ($>30^{\circ}$), with a slow and dense outflow component along the sides of the jet cone ($1-80\,\frac{km}{s}$) and a faster inner jet component ($>150\,\frac{km}{s}$). The best fits are found when the centre has a very low density.
\item The maximal outflow velocity is similar to the escape velocities of main-sequence stars and not similar to those of compact objects, such as white dwarfs. This can be seen directly from the broadness of the $H-\alpha$ absorption feature once the inclination has been determined. The resulting maximum outflow velocities correspond to a main-sequence star and not to a compact object, such as a white dwarf.
\item  The best-fit mass accretion rates are higher than expected for all of the systems. The current mass loss of the primary is too low to account for this accretion, so it is assumed that the circumbinary disk is the main source of accreted material.
  The estimated mass accretion rates are high, and using the estimated total mass of the circumbinary disk, we can estimate the lifetime of the circumbinary disk to be on the order of 100-1000 years \citep{bollen_determining_2020}. The jet phenomenon should, thus, be transient. This is in tension with the high number of jets detected in post-AGB binaries \citep{kluska_population_2022, kamath_optically_2015, manick_establishing_2017}.
\end{itemize}

The models used so far are a purely geometric description of the jets.
In this work, we report on our implementation of physical, self-similar jet models \citep{jacquemin-ide_magnetically_2019} in the same framework presented above.
We evaluate the models on the basis of the synthetic observables that imprint onto the spectroscopic time series.

\section{Inclusion of a physical jet model} \label{Self-con model}
\subsection{Accretion-ejection model}


We wish to use a magnetically driven jet model that is self-consistent with the underlying accretion disk physics.
In this accretion-ejection model, a large-scale vertical magnetic field
Bz is assumed to thread the disk.
An MHD turbulence is then triggered within the disk, which provides an anomalous (turbulent) viscosity but also magnetic diffusivity, thereby allowing accretion to proceed (see e.g. \cite{balbus_enhanced_2003} and references therein).
However, the existence of a vertical field also leads to the launching of jets, which provide another means of disk angular momentum removal \citep{blandford_hydromagnetic_1982}.
Semi-analytical models are possible thanks to the use of a self-similar Ansatz where all axisymmetric and steady-state quantities are assumed to be power laws of the cylindrical radius (see e.g. \cite{ferreira_magnetized_1995, ferreira_magnetically-driven_1997}).
This mathematical framework allows for an exact variable separation and transforms the full set of MHD steady-state partial differential equations into ordinary equations on a self-similar variable x = z/r.
The turbulent disk of isothermal scale height $h(r) = \epsilon r$, where the disk aspect ratio $\epsilon$ is used as a free constant, is then described within a resistive MHD framework.
The disk is in a magneto-hydrostatic vertical balance that is slightly sub-Keplerian accretes mostly due to the dominant jet torque.
The steady-state is then possible only if the assumed turbulence allows plasma to slip through the magnetic field within the disk.
Above the disk, turbulence is expected to be quenched, and the ejected plasma enters an ideal MHD regime.
Since bipolar jets are emitted, the disk accretion rate is a function of the radius, $\dot{M}_a(r) \varpropto r^{\xi}$, where $\xi$ is the local disk ejection efficiency.
The integration starts from the disk equatorial plane at z = 0 in the midst of the resistive MHD zone, and it proceeds in the vertical direction until the flow becomes an ideal MHD, achieves a super slow magnetosonic speed (SM critical point) above the disk surface then a super-Alfvénic poloidal speed (A critical point), and finally obtains a (conventional) super-fast magnetosonic speed.
We refer the interested reader to the work of \cite{jacquemin-ide_magnetically_2019} and references therein, which is the last published paper of a series on these exact MHD accretion-ejection solutions.\\

In this paper, we do not explore the parameter space of such MHD solutions and instead use one particular solution that has been already used in the context of Young Stellar Objects (YSOs) by \cite{panoglou_molecule_2012}.
Although it has been shown to provide slightly too fast outflows for YSOs, it has been well documented and explains most YSO jet features.
Moreover, it provides a suitable reference for our post-AGB binary system since the companion object in our case is also a solar mass star.
Such an MHD solution is described by the following set of parameters: a disk aspect ratio $\epsilon = 0.03$, a disk ejection efficiency $ \xi= 0.04$, MHD turbulence
level $\alpha_m = 2$, magnetic Prandtl number $P_m = 1$, MHD turbulence anisotropy $\chi_m = 1.42$.
We also assumed some additional heating at the disk surface so that outflows are actually magneto-thermal with a parameter f = 0.0066, corresponding to 0.66\% of the released power that goes into heating \cite{casse_magnetized_2000}.
As a consequence of a large disk magnetisation $\mu = 0.351$, accretion is supersonic within the disk (with a sonic Mach number $ m_s= 1.3$), jets carry away nearly all the angular momentum of the underlying disk, and the total jet power is about 88\% of the released accretion power, with only $ \backsim 12 \%$ of that power being radiated by the disk. We note that this energy
budget applies only to the portion of the disk (located between radii $r_{\text{in}}$ and $r_{\text{out}}$) that launches the MHD jets.
For any radius $r_o$ inside this range, the rotation velocity, density, and magnetic
field at the disk midplane are respectively:

\begin{align}
    \Omega_o r_o &= \delta \Omega_{K_o} r_o = 29.8 \delta m^{1/2} \left(\frac{r_o}{1 AU} \right)^{-1/2} km/s, \\
    n_o &= \frac{4.7}{m_s}10^{-14} m^{-1/2} \dot{m}_{-5} \left(\frac{\epsilon}{0.01} \right)^{-2} \left(\frac{r_o}{1 AU} \right)^{\xi-3/2} \left(\frac{r_{\text{in}}}{1AU} \right)^{-\xi} cm^{-3},\label{eq:density}\\
    B_z &= 2.9 m^{1/4} \dot{m}_{-5}^{1/2} \left( \frac{\mu}{m_s} \right)^{1/2} \left(\frac{r_o}{1AU} \right)^{-5/4+\xi/2} \left(\frac{r_{\text{in}}}{1AU} \right)^{-\xi/2} G,
\end{align}
where $m = M_2/M_{\odot}$ ,$\dot{m}_{-5} = \dot{M}_a(r_{in})/(10^{-5} M_{\odot} yr^{-1})$.
For the solution used here, the Keplerian deviation was found to be $\delta = 0.9$.
At the Alfvén critical point (located at $r_A/r_o = 3.8$, $z_A/r_o = 3.1$), the flow has already achieved a poloidal velocity that is 2.1 times the Keplerian velocity ($\Omega_{K_o}r_o$) at the anchoring radius $r_o$ and 1.9 times the local
rotational speed.
Asymptotically, the jet speed almost reaches its maximum velocity $v_{\text{jet}} = \sqrt{2 \lambda - 3 }\Omega_{K_o}r_o = 4.8 \Omega_{K_o}r_o$, with a
magnetic lever arm $\lambda = 1 + \frac{1}{2 \xi} \approx 13.1$.

\subsection{From line to 3D grid}

\begin{figure}
    \centering
    \includegraphics[width = 0.5\textwidth]{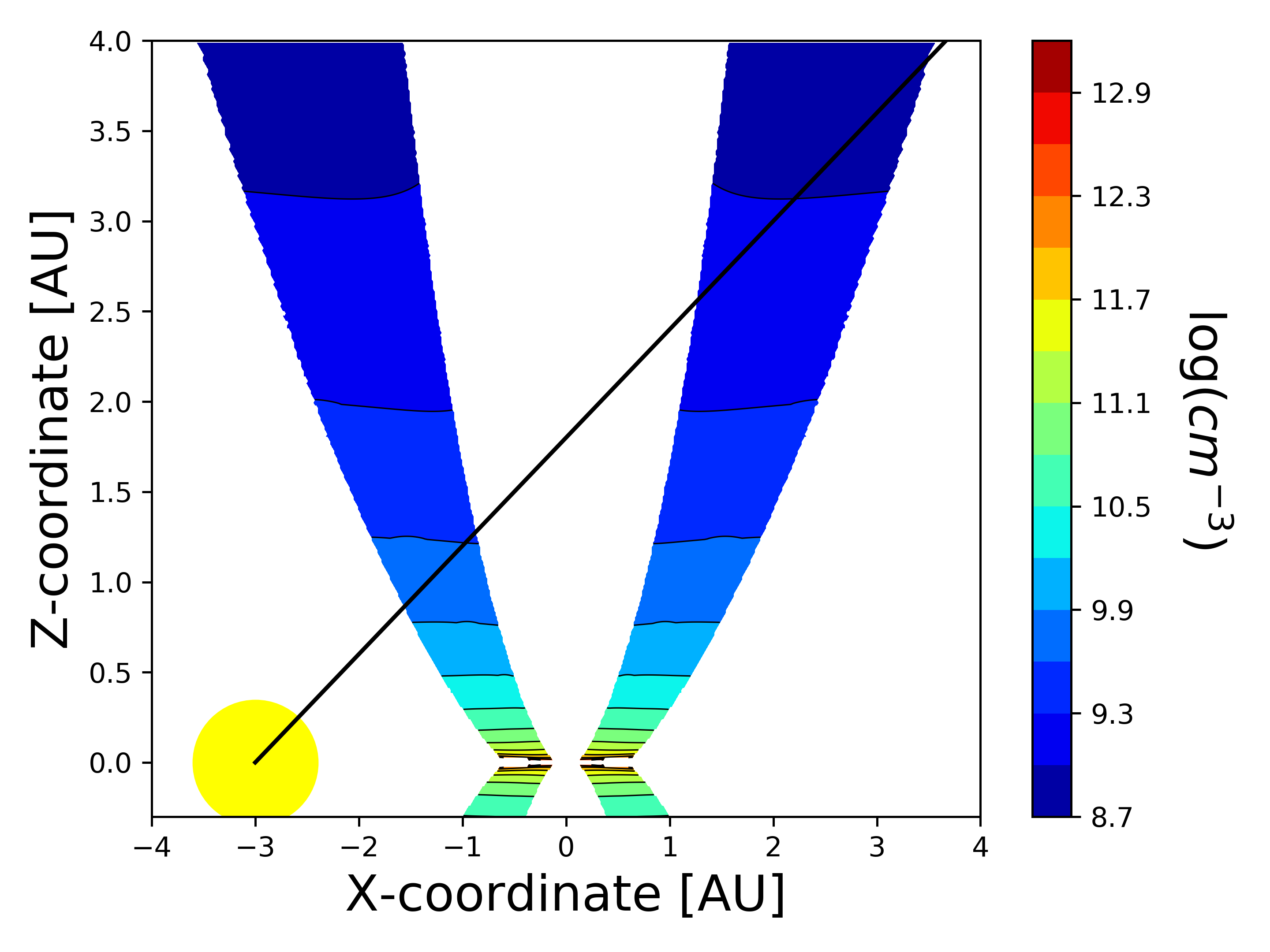}
    \caption{MHD disk wind density with LOS. Schematic of the central region of a post-AGB binary showing the post-AGB star in yellow. The jet is shown as starting from the accretion disk with the densities plotted in the colour plot. The black line shows how a hypothetical LOS would pass through the jet. For this figure as well as Figures \ref{fig:Velocity_Rot} and \ref{fig:Velocity_Vec}, the wind values are $\dot{M}(r_{in}) = 10^{-4} \frac{M_{\odot}}{yr}$, $r_{in} = 20 R_2$, and $r_{out} = 0.5 R_L$ with an inclination of $60^{\circ}$.  }
    \label{fig:LOS through jet}
\end{figure}

\begin{figure}[h]
    \centering
    \includegraphics[width = 0.8\linewidth]{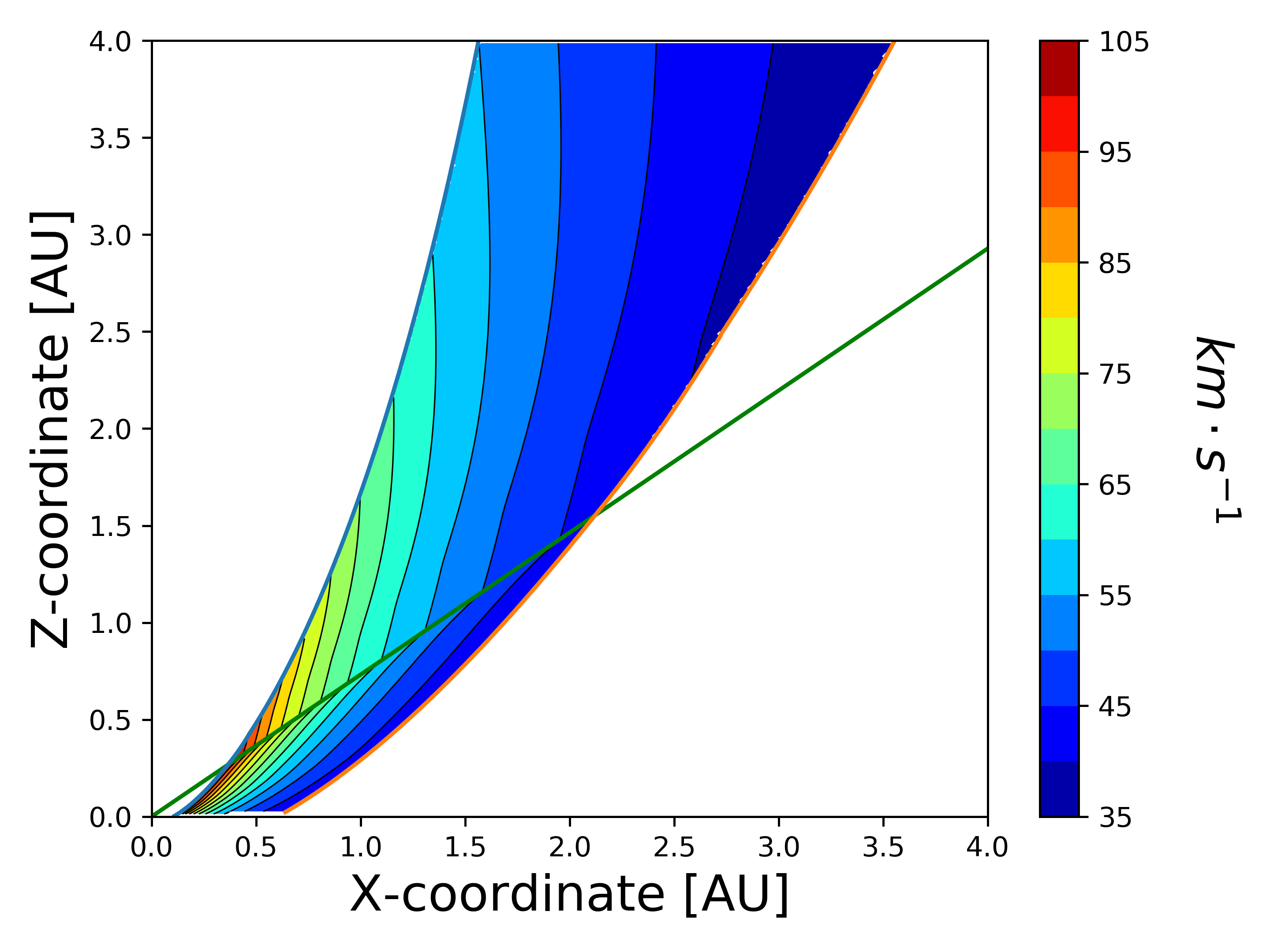}
    \caption{Rotational velocity of MHD disk wind. The figure shows an example of the rotational velocity in the jet ($v_{\phi}$); the direction of this velocity is into the page. Up to the Alfvén surface (green line), the magnetic field almost enforces isorotation to the plasma despite the opening of the flow. The decrease of the rotation velocity occurs only beyond the Alfvén surface. The wind values for this plot are the same as in Figure \ref{fig:LOS through jet}. }\label{fig:Velocity_Rot}
\end{figure}

\begin{figure}
    \centering
    \includegraphics[width = 0.8\linewidth]{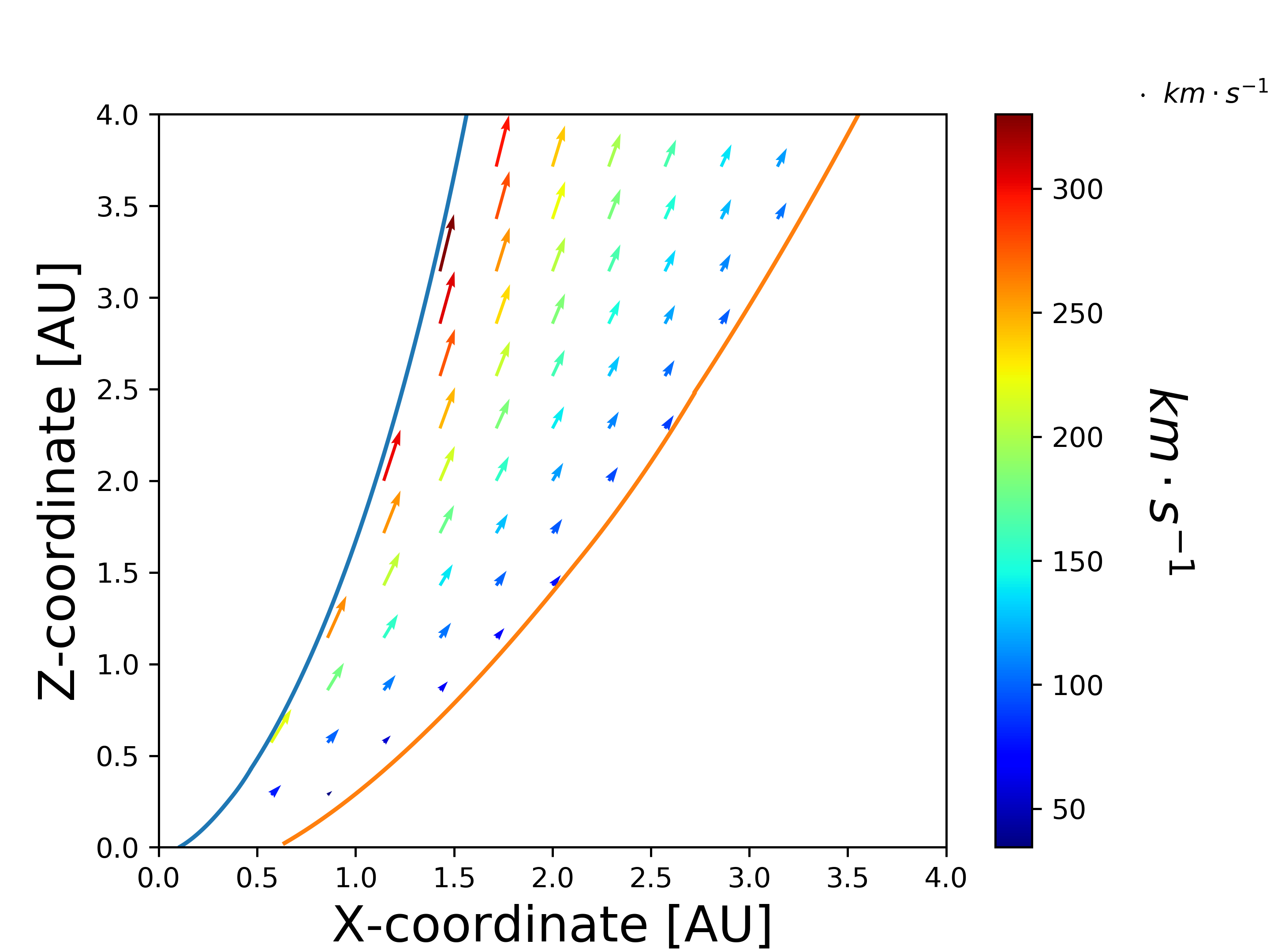}
    \caption{Velocity distribution of MHD disk wind. The figure shows an example of the velocity of a jet in the plane ($v_r$ and $v_z$). The wind values for this plot are the same as in Figure \ref{fig:LOS through jet}.}\label{fig:Velocity_Vec}

\end{figure}

Originally these models have mostly been used in single star objects.
As a result, the models have been stored in such a way that allows for the direct calculation of the required values along a LOS towards the central object of the accretion.
However, it is possible to combine the results from all possible angles and compute the density and velocity field in a grid as long as the points in the grid fall within the magnetic field lines piercing the edges of the accretion disk.
This allowed us to compute along a new LOS that is not towards the central accretor.
This is required, as the disk wind is backlit by the actual object we observe, namely, the luminous post-AGB star.
The shape, velocity, and density that are typical for these extended disk winds are shown in Figures~\ref{fig:LOS through jet}, \ref{fig:Velocity_Rot}, \ref{fig:Velocity_Vec}.
By converting this grid to a Cartesian grid, the determination of the position of each point of the jet becomes less computationally intensive.
Figures \ref{fig:Velocity_Rot} and \ref{fig:Velocity_Vec} show the Cartesian integrated velocity in the three dimensions over the jet, while Figure~\ref{fig:LOS through jet} shows the density of the jet.

From the 2D jet model described above, we produced a 3D grid by axial symmetry with the z-axis.
As the jet follows the secondary, the motion of this secondary influences the velocity field of the jet with respect to the observer.
The velocity in the models is given in three directions (radial velocity from the centre of the secondary star $v_R$, angular velocity $v_{\phi}$, and vertical velocity $v_z$).
We converted the cylindrical velocities to Cartesian, as the LOS is already set in Cartesian coordinates, using the following conversion:
\begin{align}
    v_x &= v_R \cdot \cos{\phi}+v_{\phi}\cdot \cos({90^{\circ}+\phi})\\
    v_y &= v_R \cdot \sin{\phi}+v_{\phi}\cdot \sin({90^{\circ}+\phi})\\
    v_z &= v_z.
\end{align}
Figure \ref{fig:LOS through jet} shows the new geometry that is centred on the post-AGB star instead of the main-sequence star.
Here, a typical LOS is also shown with a black line.
Outside our computed jet, the density is just set to zero, thereby neglecting any other wind source (post-AGB wind as well as any wind from the companion star).

\section{Parameter study}\label{Parameter}

In our parameter study, we studied the impact of the physical parameters of the jet model on our synthetic observables.
We used the synthetic dynamic $H_{\alpha}$ profiles of the binary post-AGB star HD\,52961 for illustration. We utilised the orbital parameters from \cite{oomen_orbital_2018} (Table \ref{tab:HD_Orbital_Param}).
We limited ourselves to the four parameters that impact the observables of a given model, which are the inner and outer launching radius, the inclination, and the mass accretion rate.

\subsection{Base model and grid}
We first defined the base model that would be modified by changing the parameters mentioned above.
The dynamic synthetic spectrum of this base model is shown in Figure \ref{fig:BaseModel}.
Since our goal was only to assess the first implications of using a physical disk wind model instead of a parametric one, we used only one MHD model, the parameters of which have been specified in Section \ref{sec:JetModel}.
The quantities that needed to be specified were (1) the innermost jet launching radius $r_{\text{in}}$, (2) the outermost jet launching radius $r_{\text{out}}$, (3) the disk accretion rate $\dot{M}_a(r_{\text{in}})$ at $r_{\text{in}}$, and (4) the LOS inclination angle i ($0^{\circ}$ is for pole-on objects, $90^{\circ}$ for edge-on).
The mass of the secondary was computed from the mass function and the inclination assuming a post-AGB star mass of $0.6 M_{\odot}$.
Additionally, the jet was set at a temperature of $5000 K$ based on the fact that the jet feature is in absorption (see the fit from \cite{bollen_jet_2021}).
The number of free parameters as a result has been reduced greatly from 12 in the parametric models to four in these new models,
with all four of these models having direct links to the physical behaviour of the jet or how we observed it.

We defined a base model with an inclination of $60^{\circ}$, a mass accretion rate of $\dot{M}_a(r_{\text{in}}) = 10^{-4} \frac{M_{\odot}}{yr}$, an inner launching radius of $20 R_2 = r_{\text{in}} = 0.09 AU$, and an outer launching radius of $r_{\text{out}} = 0.9 R_{L} = 1.6 AU$.
As the inclination sets the mass of the secondary, it also influences the stellar radius and the Roche-lobe radius.
The choice of the orbital parameters as well as the disk parameters of the base model are given in Table \ref{tab:HD_Orbital_Param}.
The choice of these parameters is based on the results obtained from the study of this object using the parametric jet model used in \cite{bollen_structure_2022} and recalled in Table \ref{tab:HD_Orbital_Param}.
In the spatiokinematic study, a high inclination as well as a very wide jet were found, which we induced by choosing a large outer launching radius.
For the inner jet launching radius $r_{\text{in}}$, we used instead a constraint on the maximum observable velocity, which we obtained along the innermost streamline.
Since it is observed to be on the order of 100 km/s (Figure \ref{fig:spectra}), we chose $r_{\text{in}} = 20R_2$ where the rotation velocity is comparable.
Nonetheless, even if the maximal jet velocity reaches $v_{\text{jet}} = 468 km/s$, the densest zone close to the disk probably dominates absorption and is associated only with poloidal projected velocities of $130-150 km/s$, in agreement with observations.\\

We subsequently undertook a parametric study of 18 more models by varying the inclination angle i, disk accretion rate, and the inner and outer jet launching radii around the mentioned fiducial values.
The inner and outer launching radii cannot be chosen completely freely. As before, $r_{\text{in}}$ has to be larger than the stellar radius, and $r_{\text{out}}$ must be smaller than the Roche lobe radius ($R_{L}$).
The ranges of parameters we explored are shown in Table \ref{tab:overview} .

\begin{table*}[htp]
    \caption{List of all the different parameters that deviate from the base model. } 
    \centering
    \begin{tabular}{c|c|cccccc}
        &base model &&&&&&\cr
        \hline
         inclination $^{\circ}$ &60&  18 & 36 & 54 & 60 & 72 & \\
         mass accretion rate [$\frac{M_{\odot}}{yr}$] & $10^{-4}$&$10^{-3}$ & $10^{-4}$ & $10^{-5}$ & $10^{-6}$ & $10^{-7}$ & \\
          outer launching radius [$R_L$] &0.9& 0.5 & 0.6 & 0.7 & 0.8 & 0.9 & 1.0 \\
          inner launching radius [$R_2$] &20& 1 & 11 & 21 & 31 & 41 & \\
          \hline
    \end{tabular}
    \begin{tablenotes}
      \small
      \item \textbf{Notes}: The base model column shows the base model from where parameters where changed. The orbital parameters of this model are shown in \ref{tab:HD_Orbital_Param}.
    \end{tablenotes}
    \label{tab:overview}
\end{table*}


\section{Results}\label{sec:results}
In this section, we first describe the results obtained with our base model. After that, we analyse the effects of the different parameters on the synthetic dynamic spectra.

\subsection{Base model}
\begin{figure}
    \centering
    \includegraphics[width = 0.5\textwidth]{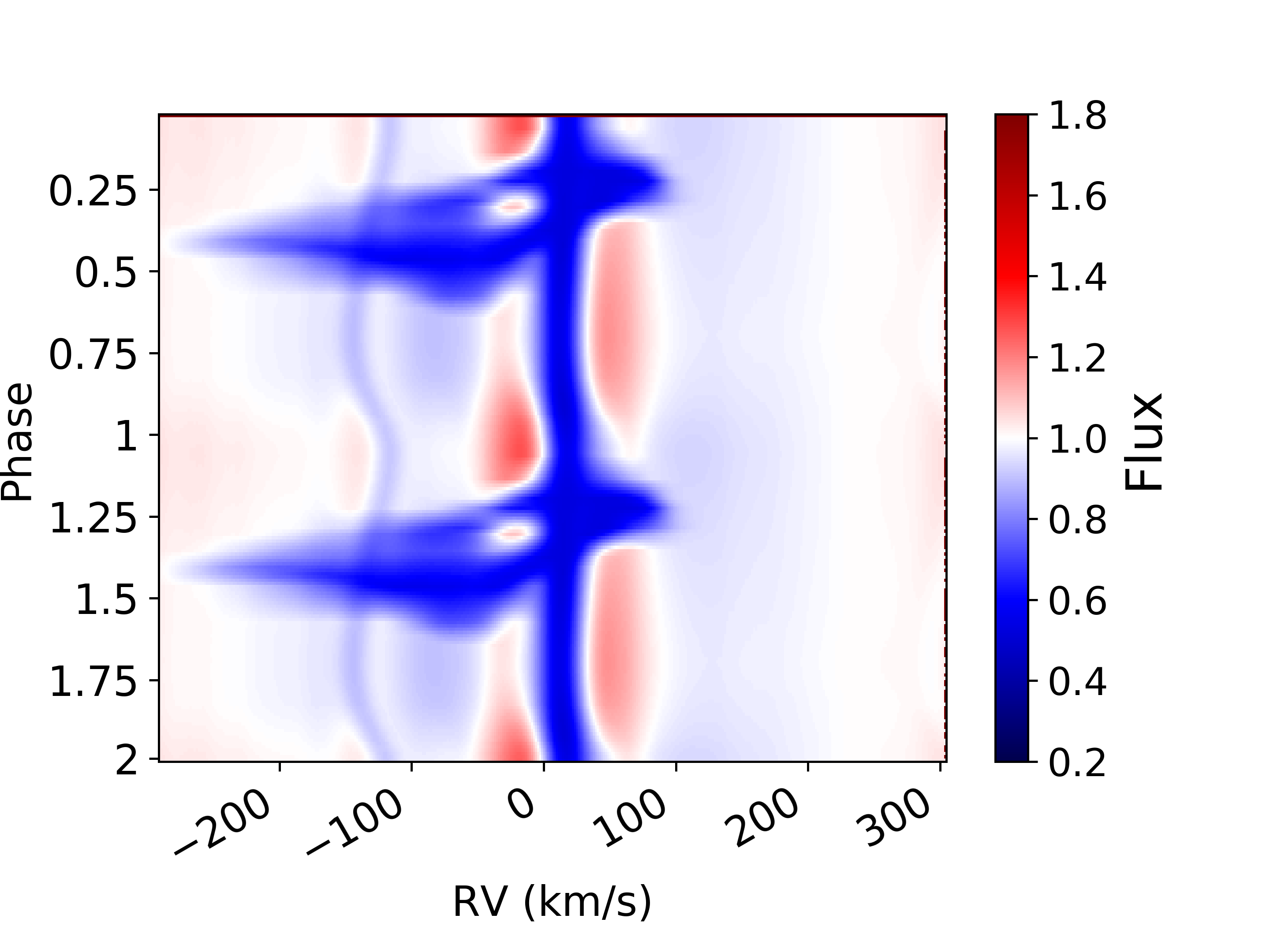}
    \caption{Base model synthetic spectra. The figure shows the synthetic dynamic spectrum using the base model. This model has an ejection efficiency of $\xi = 0.04$, a magnetisation of $\mu = 0.351$, a scale height of $\epsilon = 0.03$, and the temperature of $5000 K$.  All the other relevant parameters can be found in Tables \ref{tab:HD_Orbital_Param} and \ref{tab:overview}.}
    \label{fig:BaseModel}
\end{figure}

The base model (see Figure~\ref{fig:BaseModel}) shows a broad absorption feature similar to what is observed in Figure~\ref{fig:spectra}.
The absorption feature is present over about 35\% of the phase starting at a phase of zero.
The velocity range of the base model reaches values up to 300 $\frac{km}{s}$, which is a lot higher than the observations show.
The most remarkable difference is, however, that the synthetic spectra also show a smaller but redshifted feature at an orbital phase of about 0.3.
This feature is less wide, only visible over 10\% of the phase, and reaches velocities of only 100 $\frac{km}{s}$.
This redshifted absorption is a result of the rotation being included in these models.
At one point in the orbit, the LOS goes through one side of the disk wind that is rotating towards the post-AGB star.
This causes the absorption feature to be asymmetric as well.
There is also a lack of absorption in the middle of the broad absorption feature, which is also not observed.
This is a result of not adding a minimum density of the jet and instead choosing the region outside of the jet to be void of any material.

\subsection{Inclination}
We modelled five different inclinations ranging from almost pole-on ($i =18^{\circ}$)  to almost edge-on  ($i =72^{\circ}$).
The inclination is naturally limited in these systems, as at very high inclinations ($i>70^{\circ}$), the circumbinary disk obscures the post-AGB star, at which point, no direct light from the post-AGB primary will be seen anymore.

\begin{figure*}
   	\centering
   	\subfigure{\includegraphics[width=0.45\textwidth]{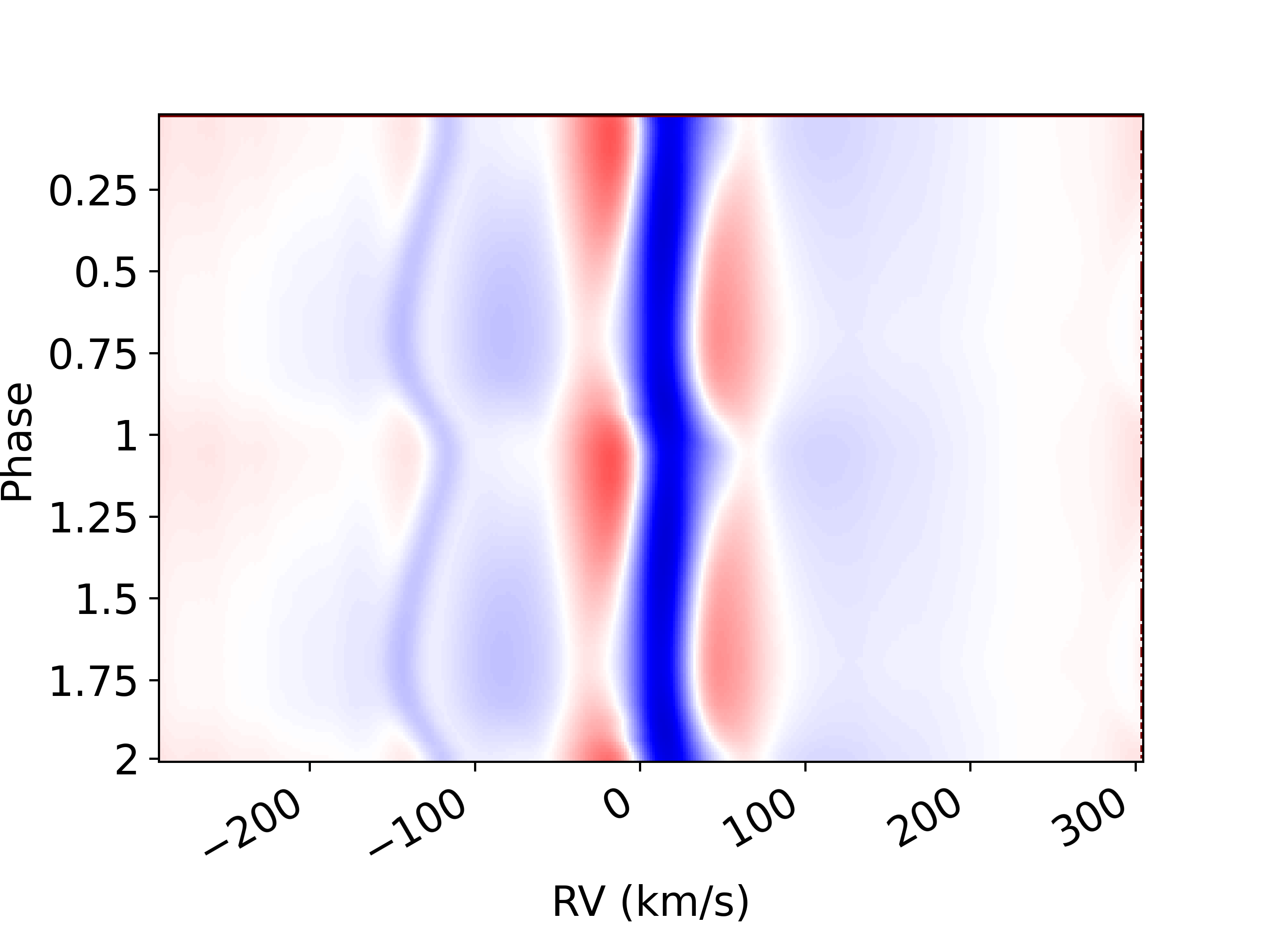}}
   	\hspace{0.025\textwidth}
   	\subfigure{\includegraphics[width=0.45\textwidth]{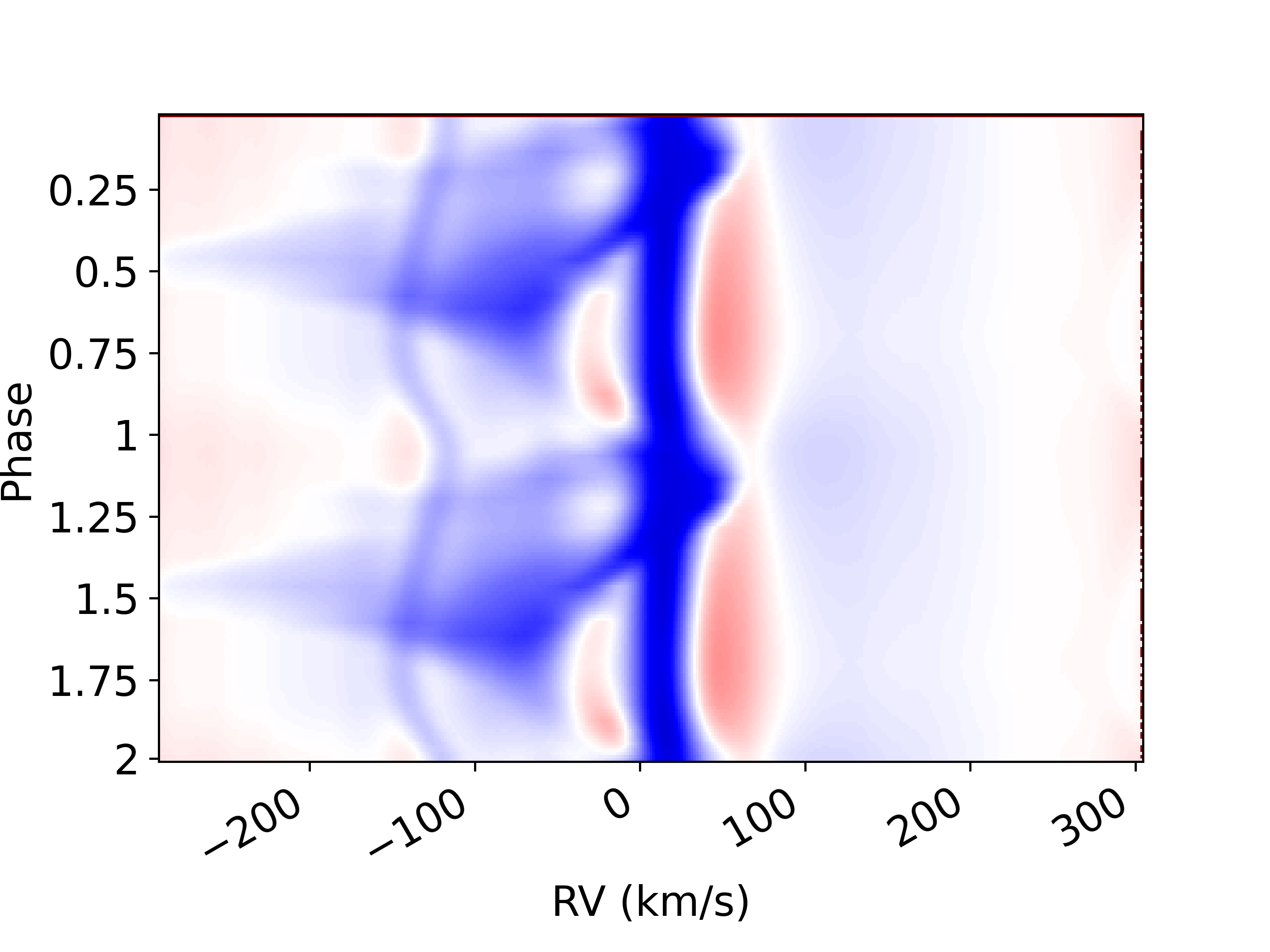}}
   	\hspace{0.025\textwidth}
   	\subfigure{\includegraphics[width=0.45\textwidth]{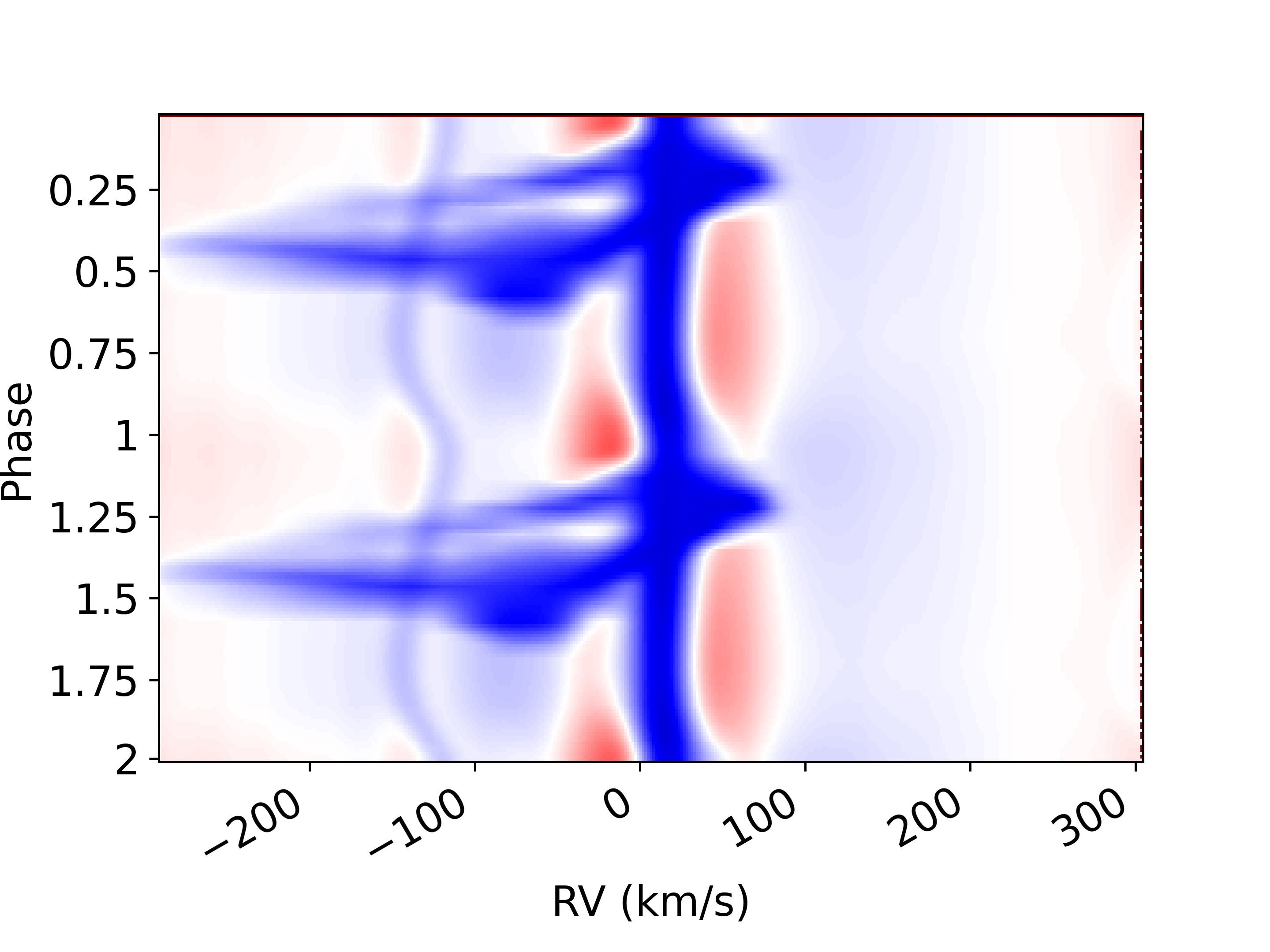}}
   	\subfigure{\includegraphics[width=0.45\textwidth]{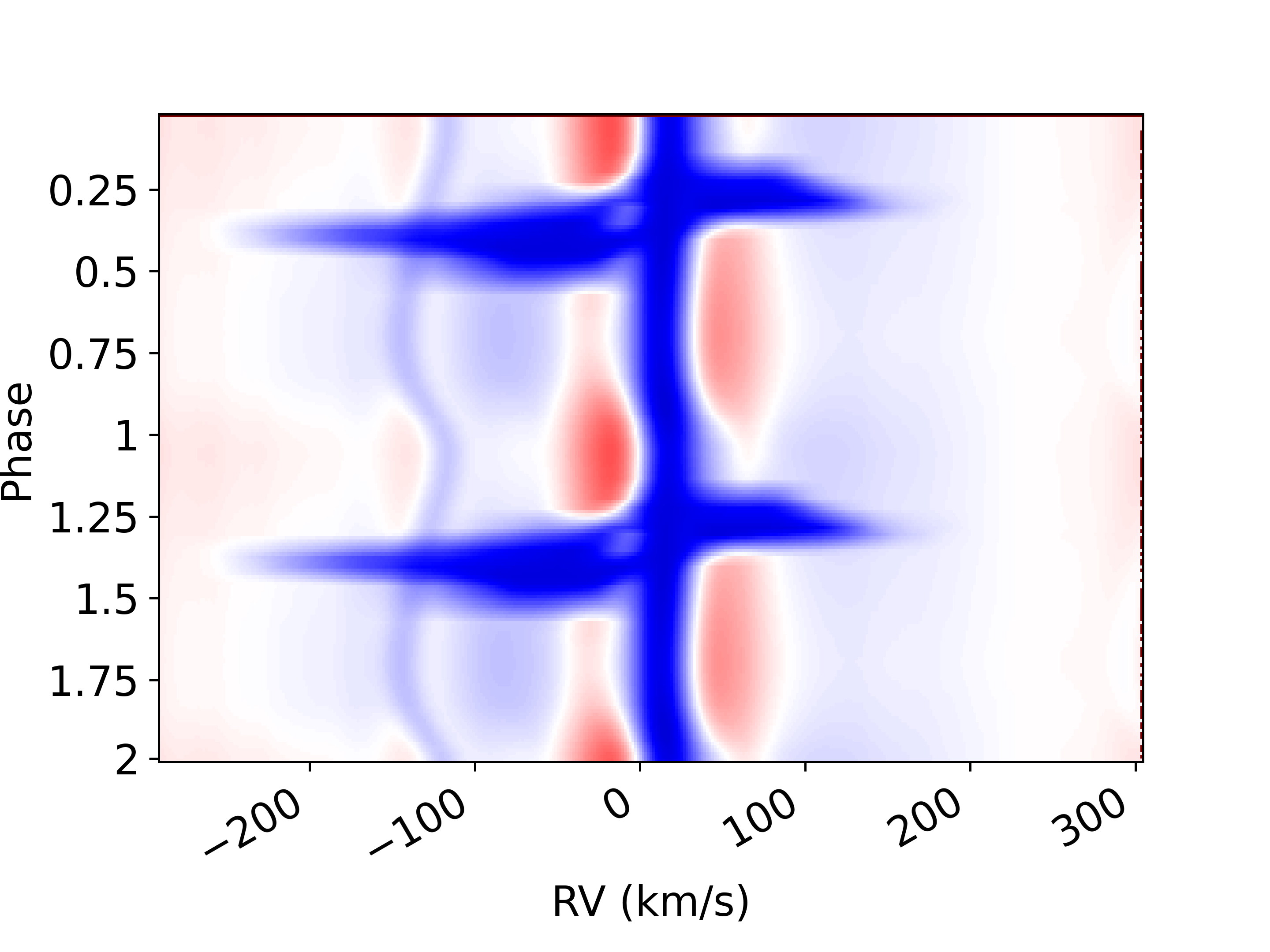}}
    \caption{Inclination effect. The dynamical spectra in this list all utilise the orbital parameters and accretion disk parameters in Table \ref{tab:HD_Orbital_Param} while changing the inclination according to Table \ref{tab:overview}. From top left to bottom right, the inclinations used are $18^{\circ}$, $36^{\circ}$, $54^{\circ}$, and $72^{\circ}$. }
    \label{fig:Inclination_Grid}
\end{figure*}

The resulting dynamic spectra are shown in Figure~\ref{fig:Inclination_Grid}.
The inclination at which we observed the system strongly influences the resulting absorption feature, as it determines through which part of the jet the LOS passes.
At lower inclinations, the LOS passes through a higher part of the jet, further from the launching region.
This results in a weaker absorption feature, as the density decreases with altitude (z-direction).
As the jet gets wider at higher altitudes, the phase-coverage of the jet becomes broader.

Additionally, at low inclinations, the rotational component of the velocity in the jet is closer to the perpendicular direction of the LOS, and the impact of this rotation to the observables disappears.
Finally, the rotational velocity is also strongest when closest to the accreting object, and by looking through higher angles, the rotational velocity decreases (Figure \ref{fig:Velocity_Rot}).

\subsection{Disk accretion rate}
\begin{figure}[h!]
   	\centering
   	\hspace{0.025\textwidth}
   	\subfigure{\includegraphics[width=0.45\textwidth]{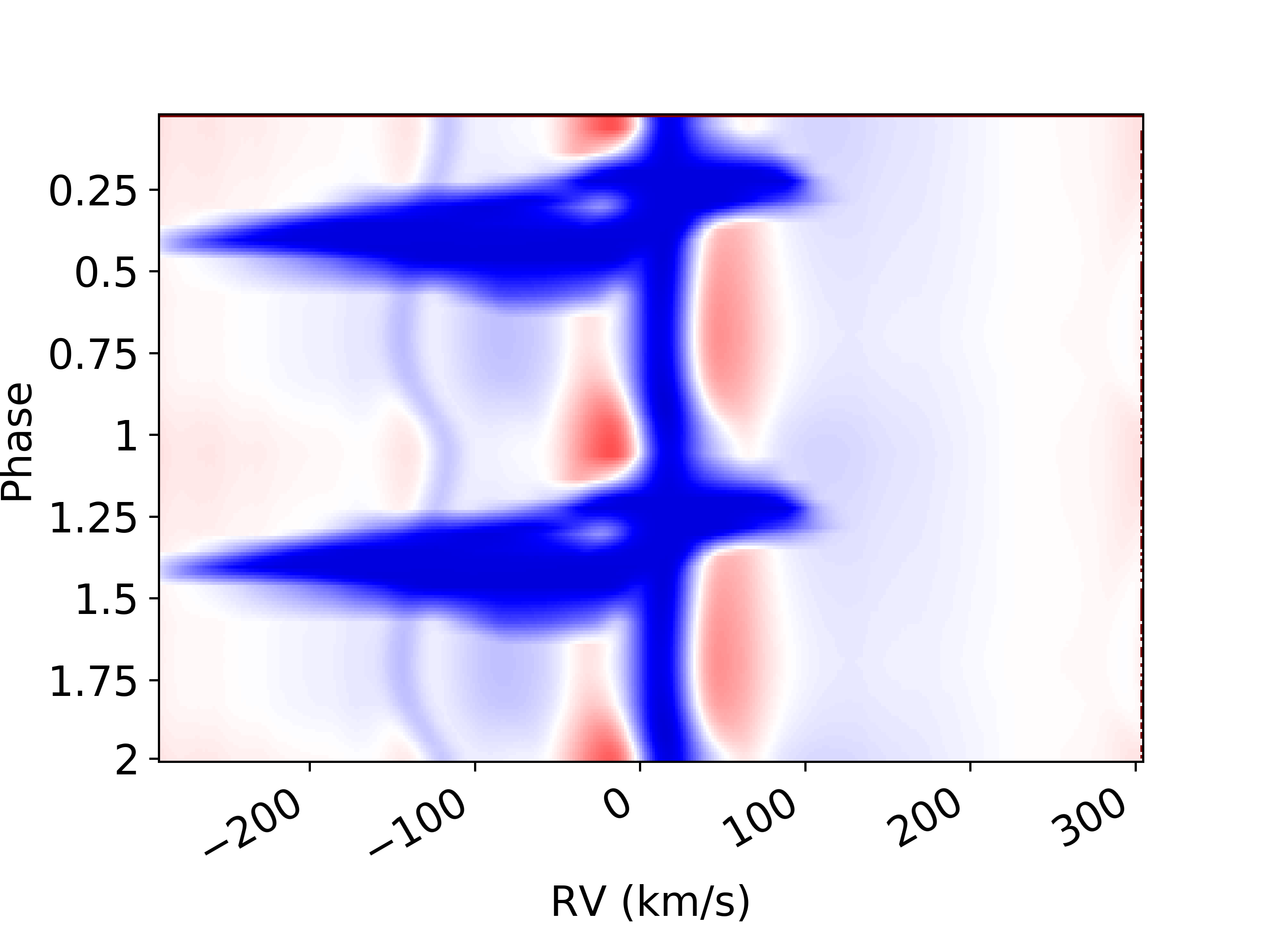}}
   	\hspace{0.025\textwidth}
   	\subfigure{\includegraphics[width=0.45\textwidth]{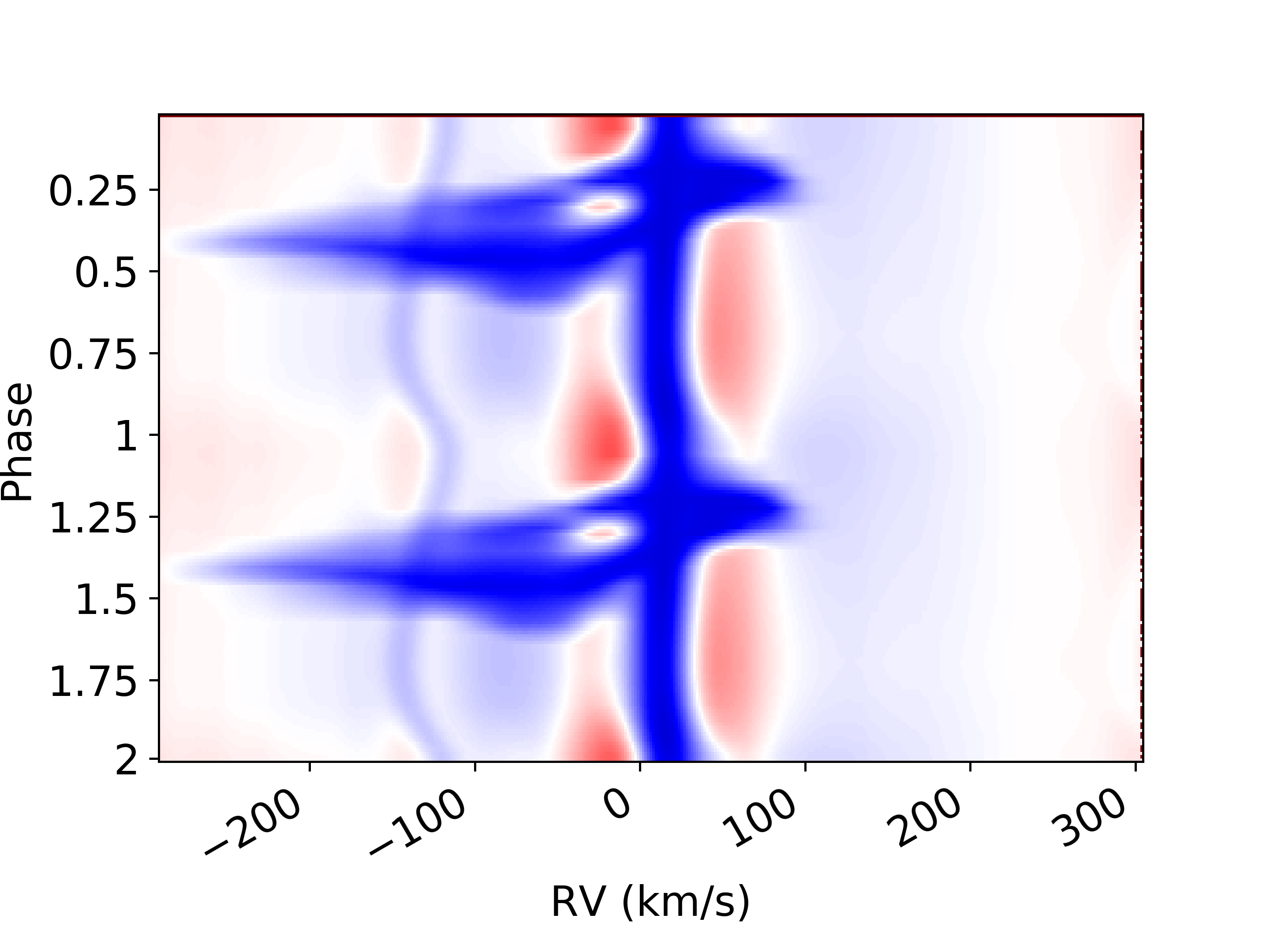}}
   	\subfigure{\includegraphics[width=0.45\textwidth]{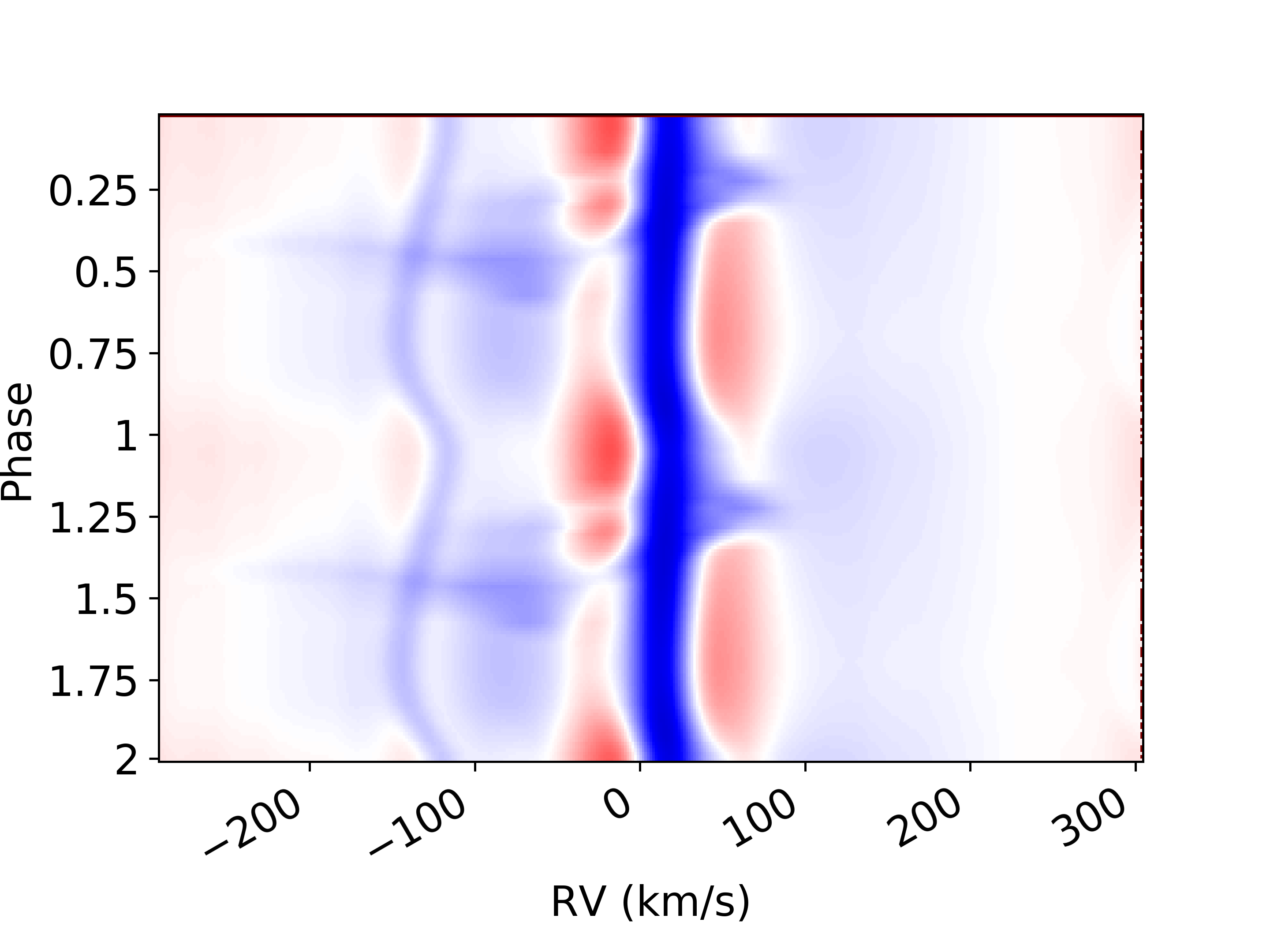}}
    \caption{Mass accretion rate effect. These dynamic spectra all have the same input parameters (Tables \ref{tab:HD_Orbital_Param} and \ref{tab:overview}) except for the mass accretion rate. These rates are, from top to bottom: $10^{-3} {M}_{\odot}/yr$, $10^{-4} {M}_{\odot}/yr$, and $10^{-5} {M}_{\odot}/yr$.}
    	\label{fig:Density_Grid}
\end{figure}

The dynamic spectra in Figure \ref{fig:Density_Grid} show that as the mass accretion is lowered,  the absorption feature becomes less strong, with no difference in the other characteristics of the dynamic spectrum.
The accretion rate directly influences the density of the jet, as can be seen in Equation \ref{eq:density}.
This is to be expected, as the disk accretion rate introduces only a scaling factor in the disk density, thereby on the ejected material.
Since the disk ejection efficiency, $\xi = 0.04$, is small, the (one-sided) mass loss in the wind is $\dot{M}_j = (\xi/2) \ln(r_{\text{out}}/r_{\text{in}}) \dot{M}_a(r_{\text{in}}) = 5.8 \cdot 10^{-6} \frac{M_{\odot}}{yr}$ for the reference model.
For the absorption feature to be noticeable, the mass accretion rate needs to be around $10^{-4} \frac{M_{\odot}}{yr}$ in the base-model setup assumed temperature.

\subsection{Launching region}
\begin{figure*}
   	\centering
   	\hspace{0.025\textwidth}
   	\subfigure{\includegraphics[width=0.45\textwidth]{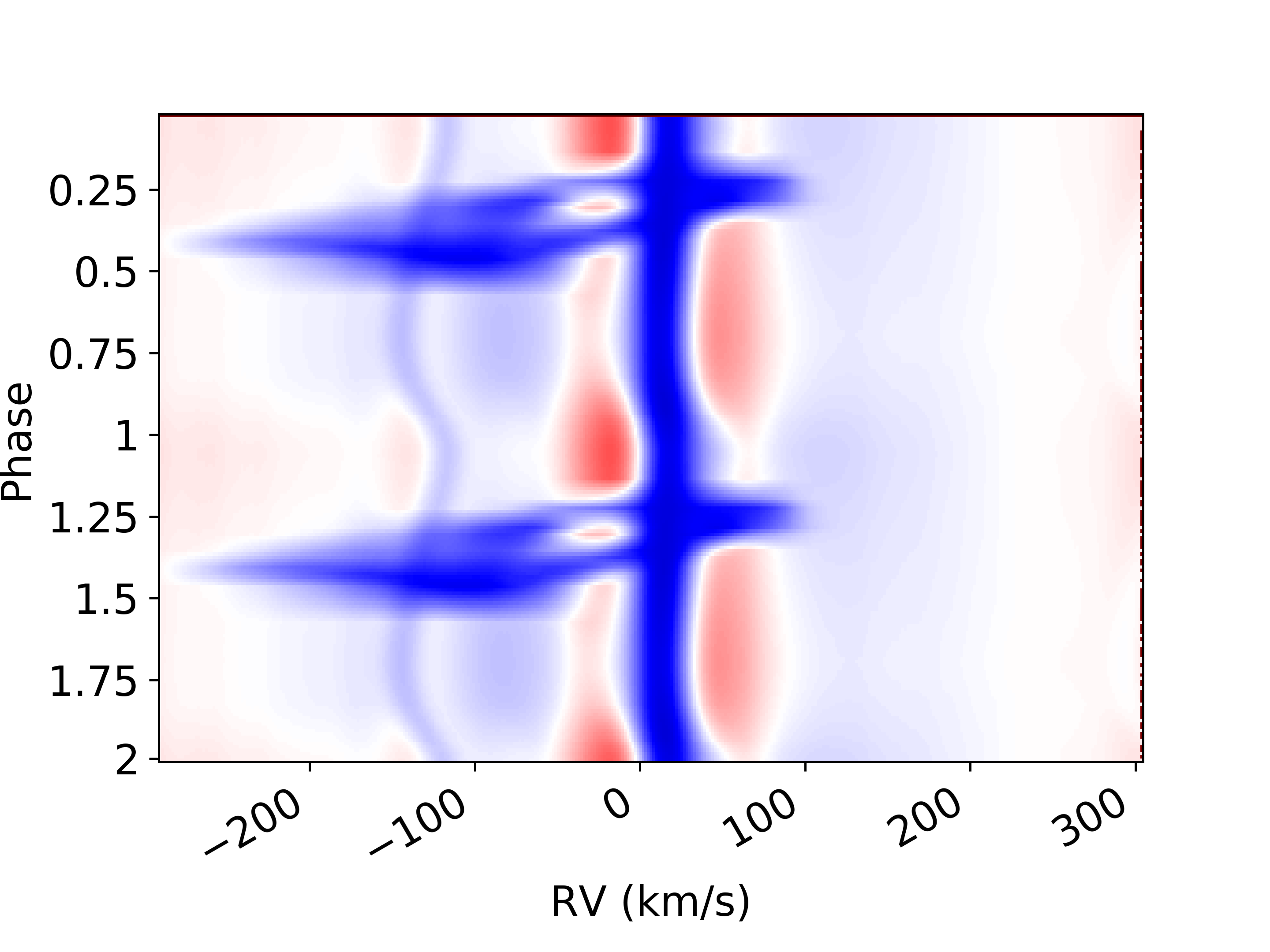}}
   	\subfigure{\includegraphics[width=0.45\textwidth]{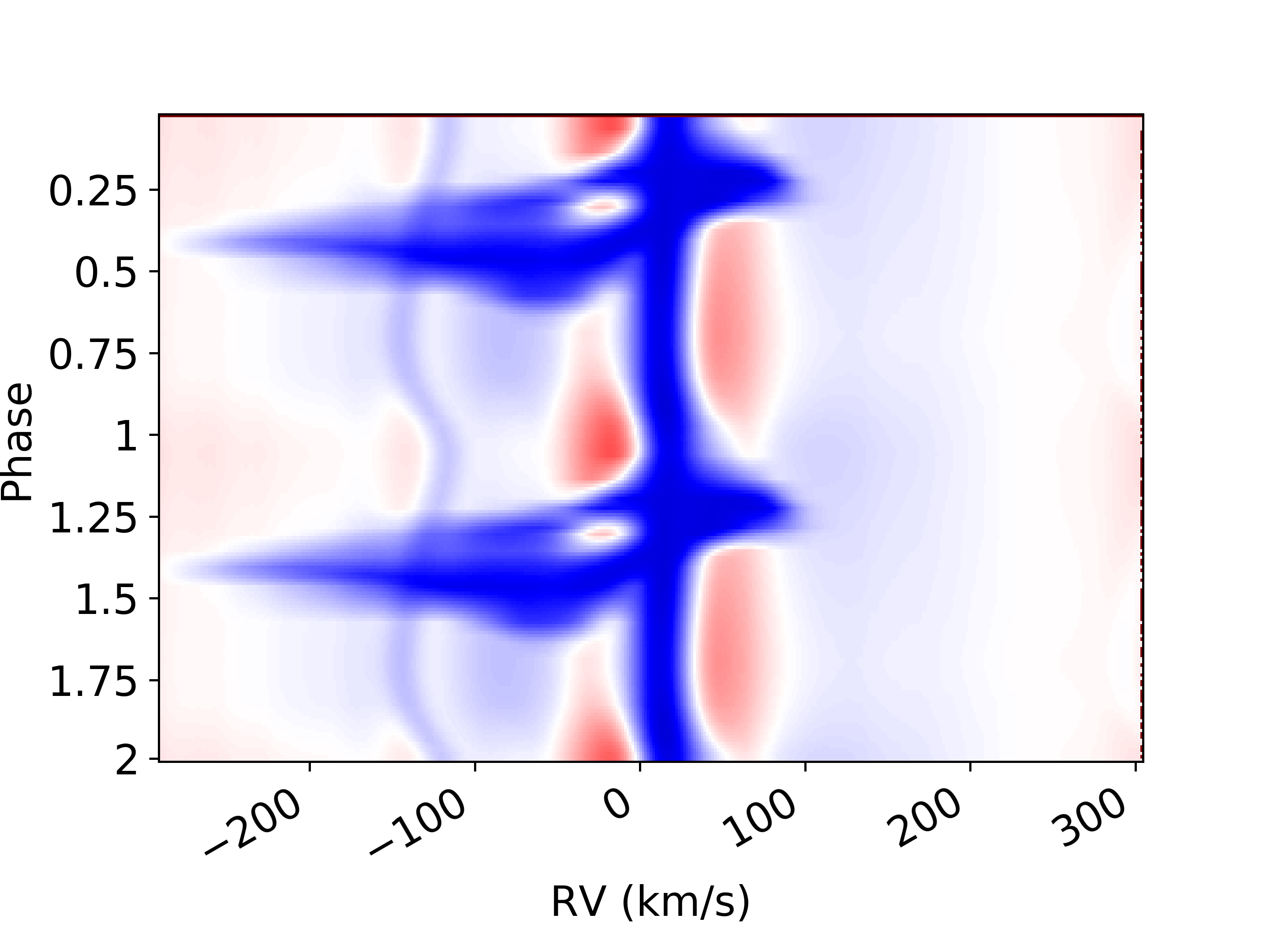}}
    \caption{Outer launching radius effect. These two dynamic spectra share all input parameters (Tables \ref{tab:HD_Orbital_Param} and \ref{tab:overview}) expect for the outer launching radius, which are, from left to right: $0.5 R_{L}$, and $1.0 R_{L}$.}
    	\label{fig:outRange_Grid}
\end{figure*}

\begin{figure}
   	\centering
   	\subfigure{\includegraphics[width=0.45\textwidth]{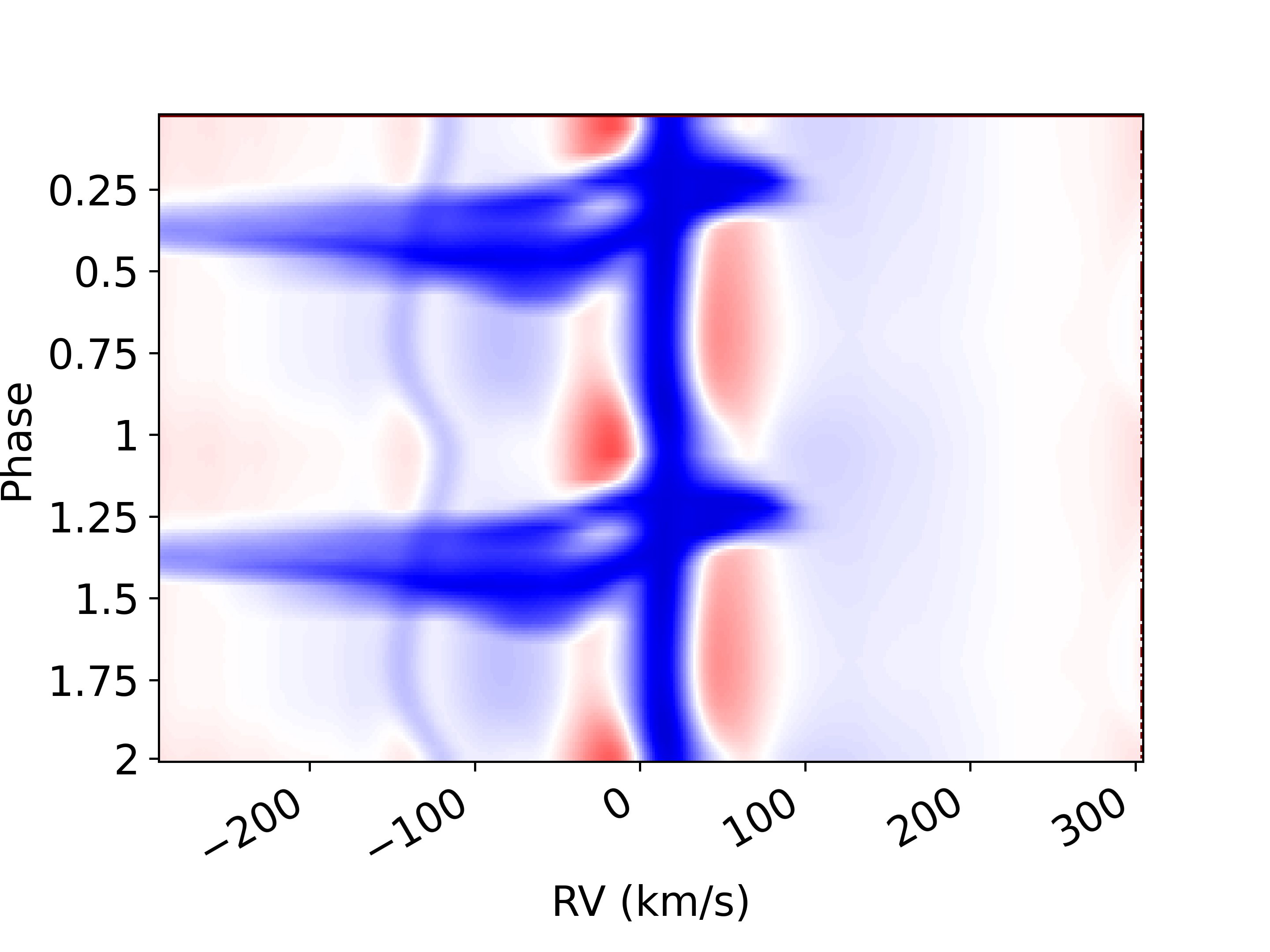}}
   	\subfigure{\includegraphics[width=0.45\textwidth]{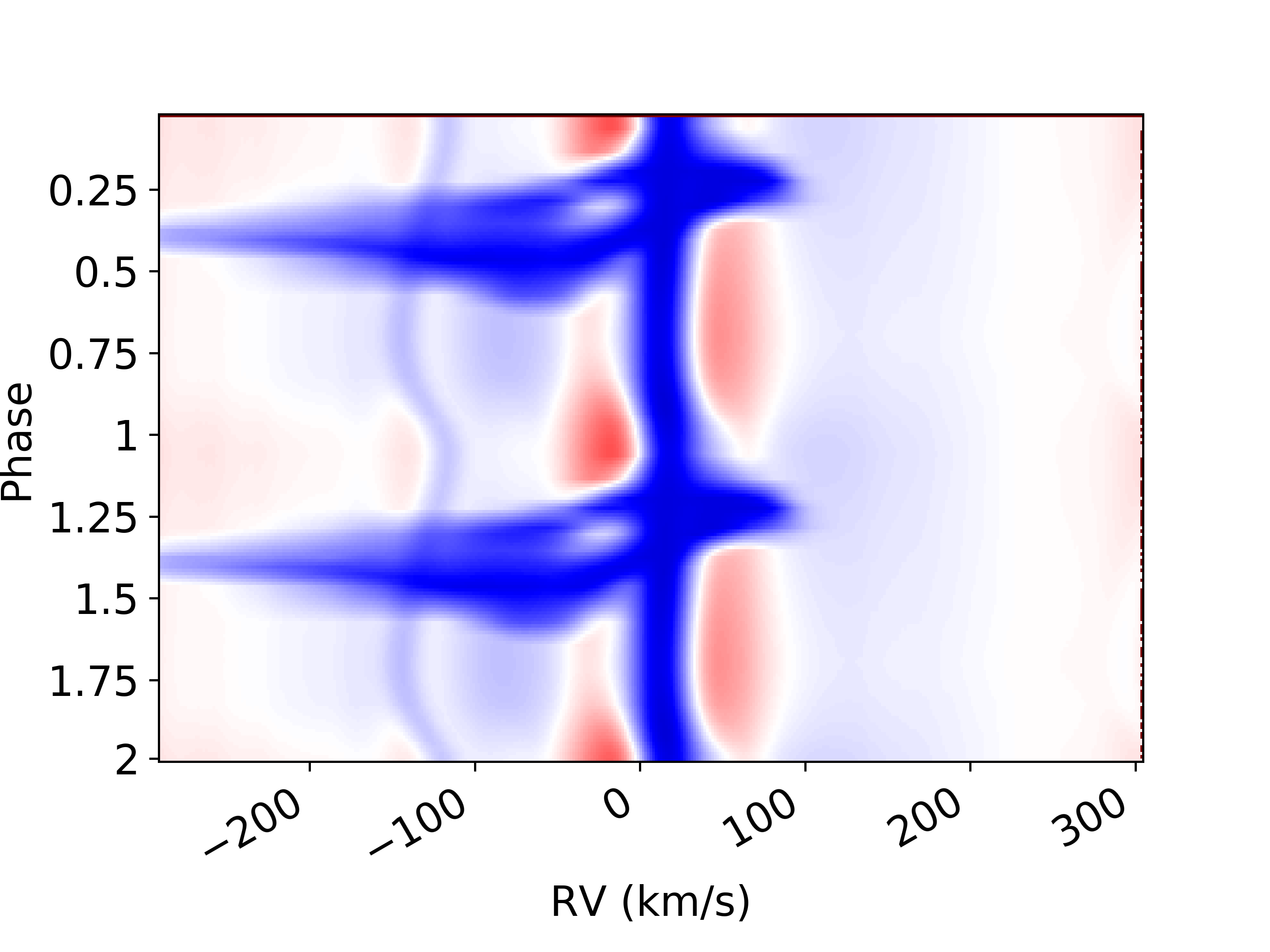}}
   	\subfigure{\includegraphics[width=0.45\textwidth]{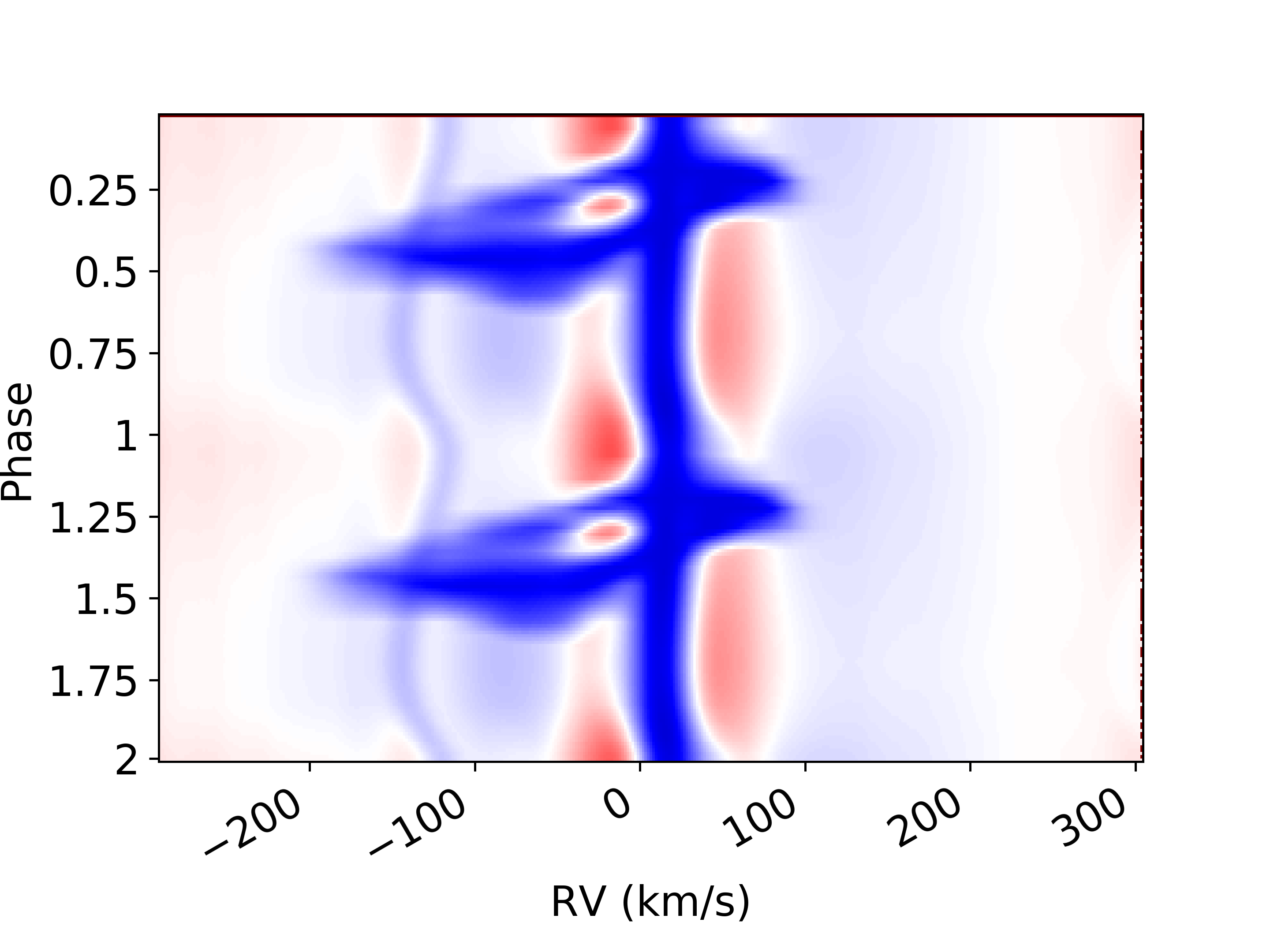}}
    \caption{Inner launching radius effect. All input parameters of these plots are the same (Tables \ref{tab:HD_Orbital_Param} and \ref{tab:overview}) excluding the inner launching radius. The values are, from top to bottom: $1 R_*$, $11 R_*$, and $41 R_*$.}
    	\label{fig:InRange_Grid}
\end{figure}

The inner and outer launching radius of the disk was changed independently for this parameter study (Figure~\ref{fig:outRange_Grid} and \ref{fig:InRange_Grid}).
Within our self-similar framework, the jet becomes wider as the launching region within the disk gets larger.
As a consequence, changing the outer jet launching radius $r_{\text{out}}$ (while keeping $r_{\text{in}}$ constant) has a strong impact on the width of the jet and therefore on the phase coverage of the absorption feature (see Figure \ref{fig:outRange_Grid}).
Increasing $r_{\text{out}}$ appears somewhat degenerate with lowering the inclination angle in the synthetic dynamic spectra.
However, and contrary to the effect of inclination, not only does the maximum velocity remain untouched when increasing $r_{\text{out}}$, but the absorption feature also becomes slightly stronger (especially at low velocities).
Figure \ref{fig:InRange_Grid} shows the influence of the innermost jet launching radius $r_{\text{in}}$ on the dynamic spectra while keeping $r_{\text{out}} = 0.9R_{L}$.
With no surprise, the most salient effect of $r_{\text{in}}$ is to change the maximum velocity seen in the absorption feature.
A low inner launching region allows the fastest part of the jet closest to the poles to be included in the jet model, which results in a larger velocity coverage at the higher end of the velocities.
On the contrary, if the inner jet launching region becomes large, the central empty core of the jet increases, and the lack of absorption present at the middle of the phase coverage
becomes larger.
This occurs because our model keeps the medium void where there is no wind.
By adding a different and additional outflow component on the symmetry axis of the bipolar jet (e.g. a stellar wind), this gap in velocity space in the synthetic spectra can be removed, but this is not included in our current model.

\subsection{Disk versus jet rotation}

\begin{figure}
    \centering
    \includegraphics[width = 0.5\textwidth]{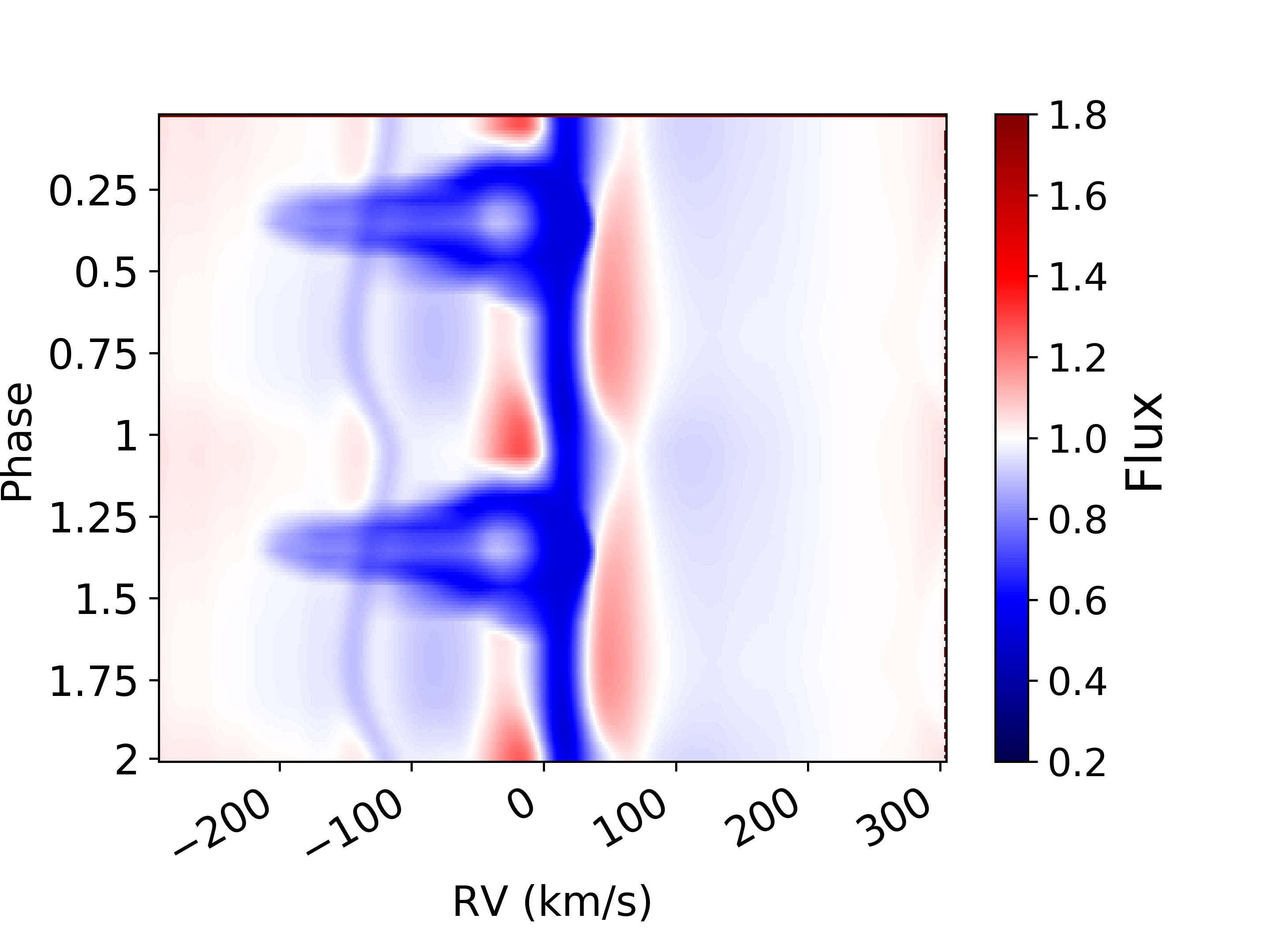}
    \caption{Synthetic dynamic spectra of a self-similar model with no rotation.}
    \label{fig:noRot}
\end{figure}

To confirm that the redshifted absorption is due to rotation, we include Figure \ref{fig:noRot}.
Here, a dynamic spectra based on a self-similar jet model for which the rotation has been set to zero and all other parameters are the same as in the base model (Figure \ref{fig:BaseModel}) is shown.
The resulting dynamic spectra indeed appear very different compared to the base model: the redshifted absorption is no longer present, and the blueshifted part is more symmetrical.
This clearly shows the importance of the rotation velocity in the production of the redshifted absorption.

\begin{figure*}
    \centering
    \subfigure{\includegraphics[width = 0.45\linewidth]{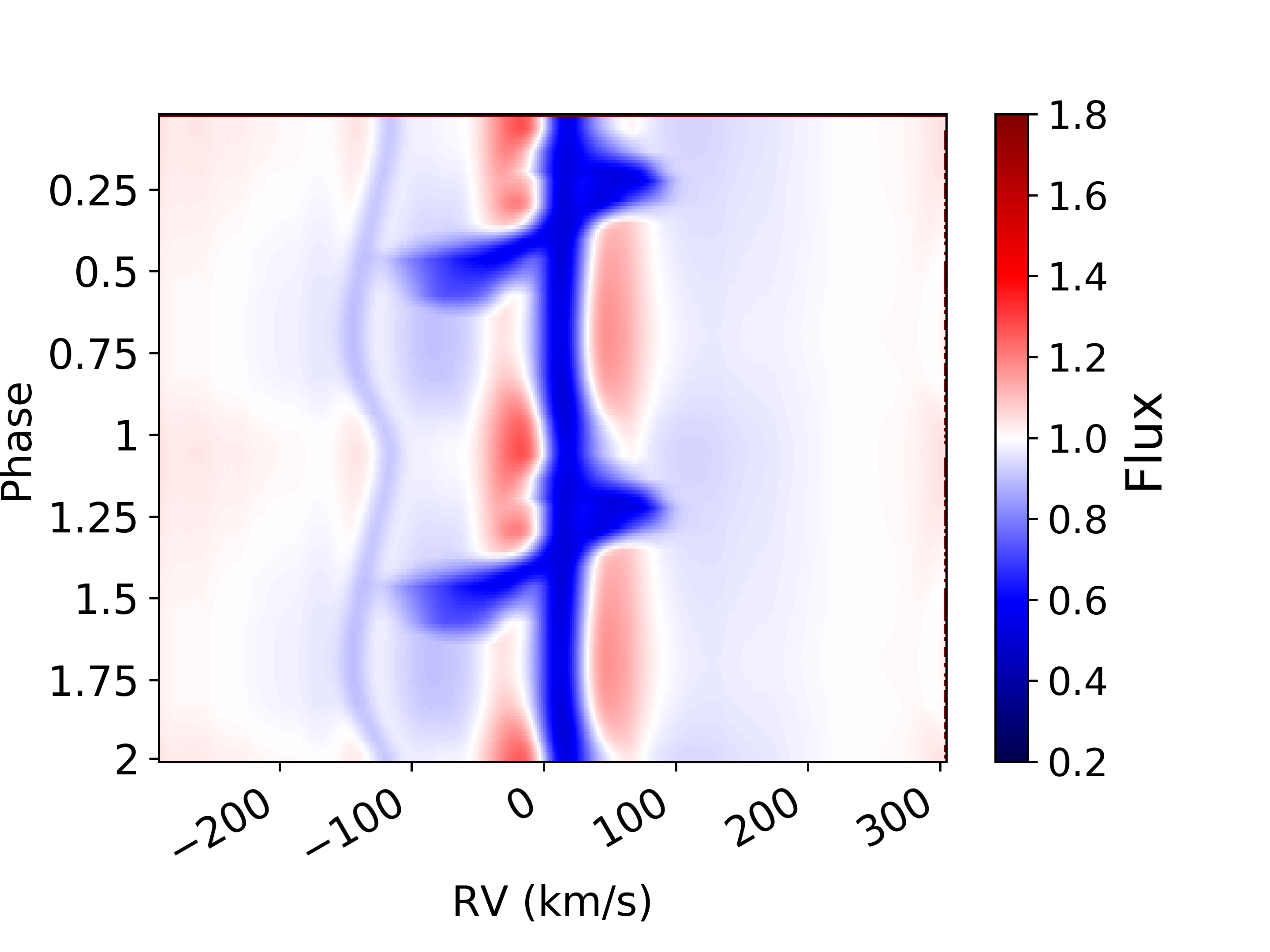}}
    \subfigure{\includegraphics[width = 0.45\linewidth]{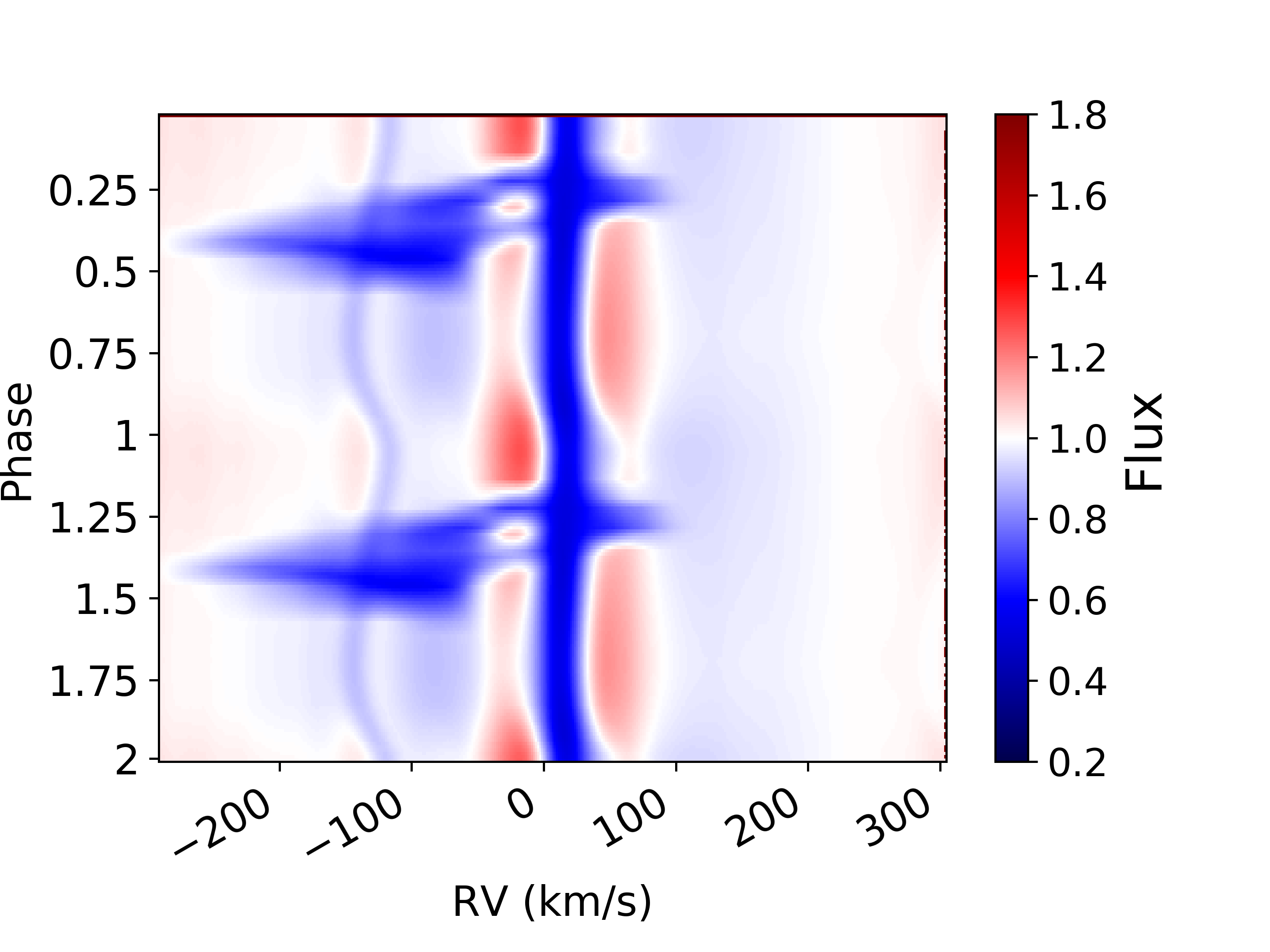}}
    \caption{Disk winds separated into sub- or super-Alfvénic. The synthetic dynamic spectra of the models only includes either sub- or super-Alfvénic flow. In the left panel, all the material that is above the Alfvénic point has been removed. In the right panel, all the material below the Alfven point has been removed.}\label{fig:Alfven_Compare}
\end{figure*}

All models with rotation produce a redshifted absorption in the spectra, something that is not observed in the data nor in the parametric models (as they do not include rotation).
To understand what part of the wind contributes to the redshifted absorption, we divided the wind in two regions: the sub-Alfvénic region and the super-Alfvénic region.
The sub-Alfvénic region includes the disk region and the starting wind, while the super-Alfvénic region is the launched wind.
We created two synthetic spectra where only the sub-Alfvénic region is included or only the super-Alfvénic region is included.
This is illustrated in Figure~\ref{fig:Alfven_Compare}.

Below the Alfven point, there is no broad blueshifted absorption present.
The redshifted absorption is comparable to the base model.
Above the Alfven point, in the jet itself, there is an extended blueshifted absorption.
The redshifted absorption is still present but appears less pronounced.
Just like the base model, there is a gap in the blueshifted part.
The region below the Alfven point has the largest contribution to the redshifted absorption component.

We made these models using an inclination of $60^{\circ}$, which is close to the highest possible inclination.
At a lower inclination, the effect of the disk rotation is less severe, which can be seen in the inclination parameter study (Figure \ref{fig:Inclination_Grid}), as the lower inclinations have less redshifted absorption.
This strengthens the idea that the regions close to the disk are the main source of the redshifted rotational feature.

\section{Discussion}

The use of a physical steady-state jet model combined with our methodology to create synthetic observables has been proven successful. The steady-state MHD disk wind model produces, for our base parameter set, velocities and phase coverage of the absorption component in the synthetic dynamic spectra in the $H_{\alpha}$ region that are very similar to our previous parametric model.
The base model, shown in Figure \ref{fig:BaseModel}, shows a reasonable accordance with the observations of HD52961 when keeping in mind that we are not fitting the observations and used only one MHD model.
However, since we are now using a physically consistent model that links the jet to the circum-companion accretion disk, several new questions arise.

An inner jet launching radius $r_{\text{in}}$ can be no smaller than $\sim 20R_2$. Otherwise, the blueshifted absorption feature would extend to velocities far larger than those observed.
This is of course related to the reason why there should be no detectable jet arising from smaller disk regions.
\\
 As with the parametric wind models, we found a rather large disk accretion rate $M_a(r_{\text{in}})$ of about $10^{-4} M_{\odot}yr^{-1}$.
If there is no way to keep wind density high while decreasing the disk accretion rate, then it raises the question as to what the lifetime of such an accretion-ejection event is.
\\
Our MHD model predicts a redshifted absorption profile due to rotation, reaching velocities of about 100 km/s. While rotation was simply ignored in the parametric jet model, it is of course to be expected from any wind emitted from a rotating body. However, in none of the 16 objects published by \cite{bollen_structure_2022}, and references therein, has a redshifted absorption feature been detected.
\\
Finally, a natural property of our MHD disk wind model is that it no longer exists inside $r_{\text{in}}$, leading to the creation of an empty core.
This is indeed one of the common features in the disk winds creating systems (e.g. \cite{bollen_structure_2022}).
However, this is also not detected in the various systems of interest.
These four items are discussed further in the following sections.
\subsection{Inner launching radius}
Within our physical framework, the companion star is surrounded by an accretion disk extending up to $ \sim R_{L}$ for simplicity and fed by the circumbinary disk.
Our main assumption is that there is a large-scale vertical field that threads the accretion disk and allows for the production of winds and jets.
Such a magnetic field is probably advected along with the ionised disk material, as in the case of X-ray binaries or cataclysmic variables.
What determines the extent of the jet emitting disk (hereafter, JED; \cite{ferreira_unified_2006}) is probably a matter of disk magnetisation, namely, whether the magnetic field is strong enough to launch jets. It sounds quite dubious to assume that such a magnetisation would decrease towards the star so that there is no clear reason why $r_{\text{in}}$ should be larger than the innermost disk radius $r_d$ itself.\footnote{We note that \cite{bollen_jet_2021} have already shown that the observed absorption features require wind velocities that are inconsistent with radiatively or thermally driven winds.}

If the companion star has a dipolar magnetic field, such a field would be coupled to the accretion disk and truncate it, enforcing the disk material to accrete along magnetospheric columns or funnels.
The position of this truncation radius thereby determines $r_d$ as is expressed in (\cite{bessolaz_accretion_2008, zanni_mhd_2009} and references therein)

\begin{align}\label{eq:trunctation}
    \frac{r_d}{R_*} &\approx 2 m_s^{2/7} \left( \frac{B_*}{140G} \right)^{4/7}  \left( \frac{\dot{M}_{a,2}}{10^{-8} {M}_{\odot}/yr}\right)^{-2/7} \left(\frac{M_2}{0.8 M_{\odot}}\right)^{-1/7}\left(\frac{R_2}{2 R_{\odot}}\right)^{5/7}\\
     &\approx 0.005 m_s^{2/7} \left(\frac{B_*}{1G}\right)^{4/7} \left(\frac{\dot{M}_{a,2}}{10^{-4} {M}_{\odot}/yr}\right)^{-2/7} \left(\frac{M_2}{1 M_{\odot}}\right)^{-1/7} \left(\frac{R_2}{ 1 R_{\odot}}\right)^{5/7},
\end{align}

\noindent
where $\dot{M}_{a,2}$ is the disk accretion rate onto the star and $B_*$ is the stellar dipole field strength (see also \cite{ireland_magnetic_2022} for numerical fits).
Since the companion star is believed to be an old main-sequence solar-mass star, we do not expect field strengths of kilo Gauss and used instead the typical $\sim 1G$ value of the Sun.
The above expression shows that for the parameters derived from our modelling, the disk would be expected to continue down to the stellar surface, namely, $r_{\text{in}} = r_d \sim R_2 = 4.6 \cdot 10^{-3} AU$.

At such a small radius, the velocity space coverage would be too large for the model used, resulting in an extremely strong redshifted absorption. This is a strong hint that this MHD model, which has been used for YSOs, does not reproduce the physical conditions around the post-AGB companion star. We note that the parametric fit also provided an $r_{in}$ much larger than $R_2$ (see table \ref{tab:HD_Orbital_Param}), so we believe this is quite generic.

\subsection{Decreasing the disk accretion rate}

The energy budget of an accretion disk undergoing mass loss through winds and jets can be written as
\begin{align}
    \label{eq:accretion power}
    P_{\text{acc}} &= 2P_{\text{jet}} +P_{\text{rad}}+P_{\text{adv},}
\end{align}
where $P_{\text{jet}}$ is the power feeding one jet, $P_{\text{rad}}$ is the disk luminosity, $P_{\text{adv}}$ is the power that is advected by the flow and released at the accretion shock onto the hard surface of the star, and
\begin{align}
    P_{\text{acc}} &= \left[ \frac{GM_*\dot{M}_a}{2r}\right]^{\text{in}}_{\text{out}} =  \frac{GM_2\dot{M}_{a, \text{in}}}{2r_{\text{in}}} - \frac{GM_2\dot{M}_{a, \text{out}}}{2r_{\text{out}}}\\
    &\simeq  \frac{GM_2\dot{M}_{a,\text{in}}}{2r_{\text{in}}} = 80 \left( \frac{\dot{M}_{a,\text{in}}}{10^{-4}{M}_{\odot}/yr} \right) \left(\frac{M_2}{1M_{\odot}} \right) \left( \frac{r_{\text{in}}}{20R_2} \right)^{-1} L_{\odot}
\end{align}
is the released accretion power.
Now $P_{\text{adv}}/ P_{\text{acc}} \simeq \epsilon^2$ and is therefore neglected, as we assumed the inner disk to be geometrically thin.

According to the jet model we used, the accretion power released up to $r_{\text{in}} = 20R_2$ amounts to $\sim 80L_{\odot}$ , 88\% of which is fed into the two jets.
However, mass within the disk must proceed down to the star with $r_d = R_2$ and at the same rate.
This will lead to an extra $\sim 1500L_{\odot}$ power, which is mostly released as radiation in the disk.
Defining the disk radiative efficiency as $\eta_{\text{rad}} = P_{\text{rad}}/ P_{\text{acc}}$ and assuming that the disk is optically thick, one gets the following rough estimate of its innermost effective temperature:
\begin{align}
    T_{\text{eff}} &= \left( \eta_{\text{rad}} \frac{G M_2 \dot{M}_a}{8 \pi \sigma r^3}\right)^{1/4}\\
    &= 3.64 \cdot 10^4 \left( \frac{\eta_{\text{rad}} \dot{M}_a(r)}{10^{-4} M_{\odot}/yr}  \right)^{1/4} \left( \frac{M_2}{M_{\odot}}\right)^{1/4} \left( \frac{r}{R_{\odot}}\right)^{-3/4} K
    \label{eq:teff}.
\end{align}
This estimate shows that an accretion disk extending down to $r_d$ around a solar-like star should have a temperature of $\eta^{1/4}_{\text{rad}}$ $36400K$.
Observationally, the contribution of the accretion disk is only a minor fraction of the total emissivity in all wavelengths.
Using high spatial resolution near-IR interferometric imaging, the contribution of the accretion disk was spatially resolved in one of the jet-producing systems (IRAS08544-4431).
This yielded a temperature of $\sim 4000 K$ only for the accretion disk in that system (\cite{hillen_interferometric_2014}, \cite{kluska_perturbed_2018}).
Thus, unless $\eta_{\text{rad}} \sim 10^{-4}$, accretion proceeding down to the star at
such high accretion rates is inconsistent with known observational constraints.
Even in the case of a JED established down to the star, and for the solution used, $\eta_{\text{rad}} = 0.12$ is a value orders of magnitude larger than required.

The other obvious solution, namely, a much smaller disk accretion rate, seems more plausible. Inverting Equation \ref{eq:teff} and assuming $T_{\text{eff}} = 4000 K$ leads to a maximum accretion rate of

\begin{align}
    \dot{M}_a (r_d) = 2 \cdot 10^{-8} \eta_{\text{rad}}^{-1} \left( \frac{T_{\text{eff}}}{4000K} \right)^{4} \left( \frac{M_2}{M_{\odot}} \right)^{-1} \left( \frac{r_d}{R_{\odot}} \right)^3 M_{\odot} yr^{-1},
\end{align}

namely, around $10^{-7} M yr^{-1}$ for a typical radiation efficiency, and a disk touching the star.\footnote{We note that inserting this expression into Equation \ref{eq:trunctation} indeed shows that unless the stellar dipole field is much larger than $1 G$, it is unable to truncate the disk, and $r_d = R_2$ is a good approximation.} This is however three orders of magnitude smaller than the value used in our base model, which is somewhat problematic.

Although they are both model dependent, our two independent constraints (wind absorption features and disk temperature) are in strong tension.
The circumbinary disk, which has a mass of approximately $10^{-2} M$ \citep{bujarrabal_detection_2015}, is believed to be the main fuel source for the circum-companion accretion disk \citep{bollen_structure_2022}. With the derived high disk-accretion rate, the lifetime would be only $\sim 10^2$ years.
If the lifetime of the circumbinary disk is this short, we would expect to find very few of these systems. In contradiction, a large number of the post-AGB binaries have a detectable disk \citep{kluska_population_2022}.

However, as seen above, a much smaller accretion rate than what used in our wind modelling would better fit the classical picture of a continuous accretion-ejection flow down to the star.
But in that case, one would need to recover approximately the same wind density as the one found in our base model.
Assuming ejection starts from $r_{\text{in}} = R_2$ ensures that the midplane density
will already be $\sim 70 - 90$ times larger than the midplane density at $20R_2$ for the same accretion rate.
But the required decrease of the disk accretion rate needs to be compensated for by a much larger disk ejection efficiency $\xi$, namely, by using another MHD model.

This necessary modification of the MHD jet model may also compensate for another consequence of putting $r_{\text{in}}$ at the stellar surface.
As shown in Figure \ref{fig:outRange_Grid}, a smaller $r_{\text{in}}$ leads to an important modification of the velocity range since at $r_{\text{in}} = R_2$, the Keplerian rotation is already 440 km/s.
The MHD wind therefore needs to provide only a mild acceleration, which is possible only with $\xi > 0.3$ or so (since $v_{\text{jet}} \simeq \sqrt{\frac{1}{\xi} - 1}\Omega_{K_o}r_o$).
But this requires absorption signatures to mostly arise from regions close to the accretion disk, where the poloidal velocity is still small.
Only a proper calculation of dynamic spectra will allow for confirmation of whether new MHD models launched close to the stellar surface can indeed reproduce the observations.

Finally, another possibility would be to better compute the jet temperature.
For simplicity, we assumed an isothermal jet, but a temperature distribution is actually expected, as a consequence of thermal and ionisation equilibrium within an expanding outflow.
By changing the jet temperature, one could, for instance, increase the occupation of the first excited state of hydrogen, which would create a stronger absorption feature at a given jet density. \\
However, the temperature of the jet is limited to the temperature of the post-AGB star (which is well determined) since the jet is always seen in absorption.
Indeed, the jet would be seen in emission instead of absorption if a higher jet temperature was
assumed compared to the effective temperature of the post-AGB star, which provides the background radiation.
Incorporating non-local thermal equilibrium (NLTE) effects in the determination of the occupation numbers of the first excited state is therefore another viable complexity increase.
This has the benefit of being able to pump hydrogen to the excitation stage without needing extra density, and by using departure coefficients, this could be done in a computationally inexpensive manner. The inclusion of NLTE effects is postponed for future work.

\subsection{The absence of rotation signatures}
The main discrepancy between our physical jet model and observations is the presence of a redshifted absorption in the synthetic spectra.
This is present in nearly all our 19 models but is missing in all dynamic spectra observed in post-AGB binaries \citep{bollen_jet_2021}.
As shown before, this is a natural consequence of a wind being emitted from a rotating body. This feature was not present in our previous investigations simply because our parametric wind model assumed zero rotation.

However, when artificially taking rotation out of our physical jet model, this redshifted feature does indeed disappear, but the absorption pattern also becomes too symmetric (see Figure \ref{fig:noRot}).
This may be a sign that some rotation is indeed present in real systems but that it is never as large as in the models.
Therefore, this puts interesting stringent constraints on wind models.
This is especially critical since, as discussed before, we expect ejection to take place up to the companion surface, where Keplerian speed will be $440 km/s$.

There are two different means of decreasing the rotation rate of an MHD wind.
The first one is to simply decrease the rotation of the disk itself so that it becomes significantly sub-Keplerian.
The deviation to the Keplerian rotation law at the disk midplane is
\begin{align}
    \delta^{2}  = \frac{\Omega_o^{2}}{\Omega^2_{K_o}} = 1- \left(\frac{5}{2} -\xi +\frac{m_s^2}{2} \right) \epsilon^2 -p \mu \epsilon,
\end{align}
where $p \sim B^+_r/B_z$ is a measure of the magnetic field curvature at the disk surface and is on the order of unity \citep{casse_magnetized_2000}.
As can be seen from this expression, rotation depends mostly on the disk aspect ratio $\epsilon$ and the disk magnetisation $\mu$.
The thicker (hotter) the disk becomes, the slower the rotation gets. In order to reach values of $\delta \sim 0.5$ or less, the disk must become geometrically thick with $\epsilon \sim 0.3$ or so.
This might, however, be problematic for purely centrifugally driven winds.
Indeed, for such a driving mechanism, the magnetic field lines need to be bent by more than $30^{\circ}$ at the disk surface (i.e. $p \geq 1$,\cite{blandford_hydromagnetic_1982}), which results in a decrease of the rotation speed of the plasma at the disk surface.
But if that surface rotation becomes too small, the centrifugal push becomes inefficient to drive ejection.
This is the reason why purely magnetically driven winds cannot be launched from thick accretion disks \citep{casse_magnetized_2000}.
However, the accretion disk in our case is irradiated by the post-AGB primary star.
This certainly provides additional heating at the disk surface so that the wind is probably also thermally driven.
The MHD solution used here already includes such a thermal effect, but it remained negligible (less than 1\%, parameter f).
Therefore, to capture the influence of the post-AGB star on the secondary accretion disk, one needs to seek thermomagnetic MHD winds with a large thermal input (f).
We note that this condition is also consistent with the constraint of having a denser wind, as
thermal effects are known to enhance the disk ejection efficiency $\xi$ \citep{casse_magnetized_2000b}.

The second way to decrease the wind rotation is by lowering the MHD efficiency of the wind acceleration.
This is done by increasing the jet mass loading parameter $\kappa$ \citep{blandford_hydromagnetic_1982}, which is related to the disk by \citep{ferreira_magnetically-driven_1997, jacquemin-ide_magnetically_2019}:
\begin{align}
    \kappa \simeq \frac{\xi m_s}{\mu} = \alpha_m p \xi \mu^{-1/2}.
\end{align}
Thus, the larger $\kappa$ is, the more mass that is loaded per magnetic flux unit, implying that the magnetic field becomes less able to shape the outflow.
It can be seen right away that the more massive the outflow (i.e. $\xi$ large) and/or the smaller the disk magnetisation (i.e. $\mu$ small), the more the outflow becomes matter
dominated.
In the large $\kappa$ limit, the magnetic lever arm $\lambda$ decreases close to unity, and the field lines become initially more conical, leading to a drastic and fast decrease of the wind rotation velocity \citep{jacquemin-ide_magnetically_2019}.

To summarise all the above constraints, a new wind model setup would require ejection down to the stellar surface $R_2$ with a disk accretion rate two or three orders of magnitude smaller than that used here, requiring an MHD wind model that is deeply modified. These modifications include: a much thicker accretion disk with $\epsilon = 0.3$ or more; a much larger disk ejection efficiency, $\xi = 0.3$ or more, probably with a significant thermal content as well (larger f value); and a lower disk magnetisation $\mu$.
This exploration will be done in a subsequent work.

\subsection{The inner axial wind}
Our disk wind model has, by construction, a central empty space between the vertical axis and the innermost field line anchored at $r_{\text{in}}$.
In order to assess the effect of only the disk wind, we arbitrarily set the density to zero in this inner region.
This void can actually be seen in the synthetic dynamical spectra as a slightly
redshifted zone of no absorption (see Figure \ref{fig:BaseModel}).
However, observations do not show any evidence of such a feature.
Moreover, our exploration of the effect of $r_{\text{in}}$ shows that even for a wind with $r_{\text{in}} = R_2$ launched near the companion star, our MHD model still displays this lack of absorption.
Again, only an approach using a different MHD model will firmly assess whether this is systematic to disk winds or dependant on the specific model used.
According to the exploration of many objects done by \cite{bollen_structure_2022}, this is systematic to disk winds.
If this is so, then it means that an extra stellar wind component must also be included in the model.
There would be no special difficulty to compute the minimum density, and hence mass loss, filling in this inner spine, which is required to recover the observed spectra.

\section{Conclusion}
A methodology to produce dynamic spectra of disk winds in post-AGB binaries using a self-consistent physical description for the wind is now functional.
This approach not only allows us to study the behaviour of the disk wind, but it also puts strong constraints on the disk itself.
In this work, we used one particular MHD model that has already been proven to explain most observational constraints in YSO jets.
The resulting synthetic spectra are able to cover velocities, phase coverages, and absorption strengths very similar to those from observations of HD\~52961.
This is very promising and indicates that magnetically driven jets launched from accretion disks may be a good representation of the actual winds detected in post-AGB systems.

However, observed velocity ranges require the jet emitting zone to be quite far $(\geq 20R_2)$ from the central companion star, which is somewhat problematic.
Moreover, observed absorption features require rather high wind densities that can only
be achieved at those distances by assuming a very large disk accretion rate (as it is self-consistent with the wind mass-loss rate).
This need for a high mass accretion rate was already found using parametric winds \citep{bollen_jet_2021} and is also problematic, as it questions the accretion disk lifetime.
Moreover, such high accretion rates should lead to the innermost disk regions being much hotter than what is currently observed.
Most remarkably, the redshifted absorption produced by the rotation of the jet does not appear in observations.

Nevertheless, we argue that the accretion disk in HD 52961 has most probably settled down to the surface of the companion star.
This is based on the conservative assumption that the stellar magnetic field is solar-like and thus unable to truncate its circumstellar accretion disk and enforce material to accrete along the magnetosphere.
Although, accretion should lead to stellar spin-up under such circumstances, we suspect that timescales are too short to lead to a revitalised dynamo action and a stellar magnetic field larger than solar-like.

As a consequence of our first use of a self-consistent accretion-ejection model, we argue that disk winds require a different MHD configuration: a larger disk ejection efficiency to
enhance wind density and lower velocities, a thicker (hotter) accretion disk to lower disk and wind rotation, and probably a smaller disk magnetisation, along with some thermal input at the disk surface (due to the post-AGB radiation field).
Such models already exist in the literature \citep{casse_magnetized_2000b, jacquemin-ide_magnetically_2019}, and the exploration of different thermo-magnetic wind models will be performed in a forthcoming paper.

\begin{acknowledgements}
OV acknowledges support from the KU Leuven Research Council (grant C16/17/007: MAESTRO).
HVW acknowledges support of the FWO under grant G097619N.
JK acknowledges support from FWO under the senior postdoctoral fellowship (1281121N).
TDP acknowledges support of the Research Foundation-Flanders (FWO) under grant 11P6I24N (Aspirant Fellowship).
DB and HVW acknowledge support from the Research Council of the KU Leuven under grant number C14/17/082.
DK acknowledges the support of the Australian Research Council (ARC) Discovery Early Career Research Award (DECRA) grant (DE190100813).
DK is also supported in part by the Australian Research Council Centre of Excellence for All Sky Astrophysics in
3 Dimensions (ASTRO 3D), through project number CE170100013.
Based on observations obtained with the HERMES spectrograph, which is supported by the Research Foundation - Flanders (FWO), Belgium, the Research Council of KU Leuven, Belgium, the Fonds National de la Recherche Scientifique (F.R.S.-FNRS), Belgium, the Royal Observatory of Belgium, the Observatoire de Genève, Switzerland and the Thüringer Landessternwarte Tautenburg, Germany.
Finally we would like to thank the anonymous referee to help streamline the paper.

\end{acknowledgements}

%
%

\addcontentsline{toc}{section}{Bibliography}
\bibliography{Bibliography/Full.bib}

\begin{thebibliography}{36}
\expandafter\ifx\csname natexlab\endcsname\relax\def\natexlab#1{#1}\fi

\bibitem[{Balbus(2003)}]{balbus_enhanced_2003}
Balbus, S.~A. 2003, Annual Review of Astronomy and Astrophysics, 41, 555

\bibitem[{Bessolaz {et~al.}(2008)Bessolaz, Zanni, Ferreira, Keppens, \&
  Bouvier}]{bessolaz_accretion_2008}
Bessolaz, N., Zanni, C., Ferreira, J., Keppens, R., \& Bouvier, J. 2008,
  Astronomy \& Astrophysics, 478, 155

\bibitem[{Blandford \& Payne(1982)}]{blandford_hydromagnetic_1982}
Blandford, R.~D. \& Payne, D.~G. 1982, Monthly Notices of the Royal
  Astronomical Society, 199, 883

\bibitem[{Bollen(2021)}]{bollen_jet_2021}
Bollen, D. 2021, PhD thesis,
  \href{https://fys.kuleuven.be/ster/pub/phd-thesis-dylan-bollen/phd-thesis-dylan-bollen}{https://fys.kuleuven.be/ster/pub/phd-thesis-dylan-bollen/phd-thesis-dylan-bollen}

\bibitem[{{Bollen} {et~al.}(2020){Bollen}, {Kamath}, {De Marco}, {Van Winckel},
  \& {Wardle}}]{bollen_determining_2020}
{Bollen}, D., {Kamath}, D., {De Marco}, O., {Van Winckel}, H., \& {Wardle}, M.
  2020, Astronomy \& Astrophysics, 641, A175

\bibitem[{Bollen {et~al.}(2019)Bollen, Kamath, Van~Winckel, \&
  De~Marco}]{bollen_spatio-kinematic_2019}
Bollen, D., Kamath, D., Van~Winckel, H., \& De~Marco, O. 2019, Astronomy and
  Astrophysics, 631, A53

\bibitem[{{Bollen} {et~al.}(2022){Bollen}, {Kamath}, {Van Winckel}, {De Marco},
  {Verhamme}, {Kluska}, \& {Wardle}}]{bollen_structure_2022}
{Bollen}, D., {Kamath}, D., {Van Winckel}, H., {et~al.} 2022, Astronomy \&
  Astrophysics, 666, A40

\bibitem[{Bollen {et~al.}(2017)Bollen, Van~Winckel, \&
  Kamath}]{bollen_jet_2017}
Bollen, D., Van~Winckel, H., \& Kamath, D. 2017, Astronomy and Astrophysics,
  607, A60

\bibitem[{Bujarrabal {et~al.}(2015)Bujarrabal, Castro-Carrizo, Alcolea, \&
  Van~Winckel}]{bujarrabal_detection_2015}
Bujarrabal, V., Castro-Carrizo, A., Alcolea, J., \& Van~Winckel, H. 2015,
  Astronomy and Astrophysics, 575, L7

\bibitem[{Casse \& Ferreira(2000{\natexlab{a}})}]{casse_magnetized_2000}
Casse, F. \& Ferreira, J. 2000{\natexlab{a}}, Astronomy and Astrophysics, 353,
  1115

\bibitem[{Casse \& Ferreira(2000{\natexlab{b}})}]{casse_magnetized_2000b}
Casse, F. \& Ferreira, J. 2000{\natexlab{b}}, Astronomy and Astrophysics, 361,
  1178

\bibitem[{Demircan \& Kahraman(1991)}]{demircan_stellar_1991}
Demircan, O. \& Kahraman, G. 1991, Astrophysics and Space Science, 181, 313

\bibitem[{Eggleton(1983)}]{eggleton_approximations_1983}
Eggleton, P.~P. 1983, The Astrophysical Journal, 268, 368

\bibitem[{Ferreira(1997)}]{ferreira_magnetically-driven_1997}
Ferreira, J. 1997, Astronomy and Astrophysics, 319, 340

\bibitem[{Ferreira \& Casse(2013)}]{ferreira_fan-shaped_2013}
Ferreira, J. \& Casse, F. 2013, Monthly Notices of the Royal Astronomical
  Society, 428, 307

\bibitem[{Ferreira \& Pelletier(1995)}]{ferreira_magnetized_1995}
Ferreira, J. \& Pelletier, G. 1995, Astronomy and Astrophysics, 295, 807

\bibitem[{Ferreira {et~al.}(2006)Ferreira, Petrucci, Henri, Saugé, \&
  Pelletier}]{ferreira_unified_2006}
Ferreira, J., Petrucci, P.~O., Henri, G., Saugé, L., \& Pelletier, G. 2006,
  Astronomy and Astrophysics, 447, 813

\bibitem[{Gorlova {et~al.}(2012)Gorlova, Van~Winckel, Gielen, Raskin, Prins,
  Pessemier, Waelkens, Frémat, Hensberge, Dumortier, Jorissen, \&
  Van~Eck}]{gorlova_time-resolved_2012}
Gorlova, N., Van~Winckel, H., Gielen, C., {et~al.} 2012, Astronomy and
  Astrophysics, 542, A27

\bibitem[{Hillen {et~al.}(2016)Hillen, Kluska, Bouquin, Winckel, Berger,
  Kamath, \& Bujarrabal}]{hillen_imaging_2016}
Hillen, M., Kluska, J., Bouquin, J.-B.~L., {et~al.} 2016, Astronomy \&
  Astrophysics, 588, L1

\bibitem[{Hillen {et~al.}(2014)Hillen, Menu, Van~Winckel, Min, Gielen, Wevers,
  Mulders, Regibo, \& Verhoelst}]{hillen_interferometric_2014}
Hillen, M., Menu, J., Van~Winckel, H., {et~al.} 2014, Astronomy and
  Astrophysics, 568, A12

\bibitem[{Ireland {et~al.}(2022)Ireland, Matt, \&
  Zanni}]{ireland_magnetic_2022}
Ireland, L.~G., Matt, S.~P., \& Zanni, C. 2022, The Astrophysical Journal, 929,
  65

\bibitem[{Jacquemin-Ide {et~al.}(2019)Jacquemin-Ide, Ferreira, \&
  Lesur}]{jacquemin-ide_magnetically_2019}
Jacquemin-Ide, J., Ferreira, J., \& Lesur, G. 2019, Monthly Notices of the
  Royal Astronomical Society, 490, 3112

\bibitem[{{Jacquemin-Ide} {et~al.}(2021){Jacquemin-Ide}, {Lesur}, \&
  {Ferreira}}]{jacquemin-ide_magnetic_2020}
{Jacquemin-Ide}, J., {Lesur}, G., \& {Ferreira}, J. 2021, Astronomy \&
  Astrophysics, 647, A192

\bibitem[{Kamath {et~al.}(2015)Kamath, Wood, \&
  Van~Winckel}]{kamath_optically_2015}
Kamath, D., Wood, P.~R., \& Van~Winckel, H. 2015, Monthly Notices of the Royal
  Astronomical Society, 454, 1468

\bibitem[{Kluska {et~al.}(2018)Kluska, Hillen, Van~Winckel, Manick, Min,
  Regibo, \& Royer}]{kluska_perturbed_2018}
Kluska, J., Hillen, M., Van~Winckel, H., {et~al.} 2018, Astronomy \&
  Astrophysics, 616, A153

\bibitem[{Kluska {et~al.}(2022)Kluska, Winckel, Coppée, Oomen, Dsilva, Kamath,
  Bujarrabal, \& Min}]{kluska_population_2022}
Kluska, J., Winckel, H.~V., Coppée, Q., {et~al.} 2022, Astronomy \&
  Astrophysics, 658, A36

\bibitem[{Manick {et~al.}(2017)Manick, Van~Winckel, Kamath, Hillen, \&
  Escorza}]{manick_establishing_2017}
Manick, R., Van~Winckel, H., Kamath, D., Hillen, M., \& Escorza, A. 2017,
  Astronomy \& Astrophysics, 597, A129

\bibitem[{Oomen {et~al.}(2019)Oomen, Van~Winckel, Pols, \&
  Nelemans}]{oomen_modelling_2019}
Oomen, G.-M., Van~Winckel, H., Pols, O., \& Nelemans, G. 2019, Astronomy and
  Astrophysics, 629, A49

\bibitem[{Oomen {et~al.}(2018)Oomen, Winckel, Pols, Nelemans, Escorza, Manick,
  Kamath, \& Waelkens}]{oomen_orbital_2018}
Oomen, G.-M., Winckel, H.~V., Pols, O., {et~al.} 2018, Astronomy \&
  Astrophysics, 620, A85

\bibitem[{Panoglou {et~al.}(2012)Panoglou, Cabrit, Forêts, Garcia, Ferreira,
  \& Casse}]{panoglou_molecule_2012}
Panoglou, D., Cabrit, S., Forêts, G. P.~d., {et~al.} 2012, Astronomy \&
  Astrophysics, 538, A2

\bibitem[{Raskin(2011)}]{raskin_hermes_2011}
Raskin, G. 2011, PhD thesis,
  \href{https://fys.kuleuven.be/ster/pub/phd-thesis-gert-raskin/phd-thesis-gert-raskin-1}{https://fys.kuleuven.be/ster/pub/phd-thesis-gert-raskin/phd-thesis-gert-raskin-1}

\bibitem[{{Raskin} {et~al.}(2011){Raskin}, {van Winckel}, {Hensberge},
  {Jorissen}, {Lehmann}, {Waelkens}, {Avila}, {de Cuyper}, {Degroote},
  {Dubosson}, {Dumortier}, {Fr{\'e}mat}, {Laux}, {Michaud}, {Morren}, {Perez
  Padilla}, {Pessemier}, {Prins}, {Smolders}, {van Eck}, \&
  {Winkler}}]{raskin_2011}
{Raskin}, G., {van Winckel}, H., {Hensberge}, H., {et~al.} 2011, Astronomy \&
  Astrophysics, 526, A69

\bibitem[{Van~Winckel(2003)}]{van_winckel_post-agb_2003}
Van~Winckel, H. 2003, Annual Review of Astronomy and Astrophysics, 41, 391

\bibitem[{Van~Winckel(2007)}]{van_winckel_post-agb_2007}
Van~Winckel, H. 2007, Baltic Astronomy, 16, 112

\bibitem[{Van~Winckel(2018)}]{van_winckel_binary_2018}
Van~Winckel, H. 2018, in: "The impact of Binary Stars of Stellar evolution",
  eds.: G. Beccari, H.M.J. Boffin, Cambridge Astrophysics Series, 54, 92

\bibitem[{Zanni \& Ferreira(2009)}]{zanni_mhd_2009}
Zanni, C. \& Ferreira, J. 2009, Astronomy and Astrophysics, 508, 1117

\end{thebibliography}
\end{document}